\documentclass[superscriptaddress,prd,aps,preprintnumbers,amsmath,amssymb,nofootinbib,preprint]{revtex4} 
\usepackage[pdftex]{graphicx}
\usepackage{dcolumn}
\usepackage{bm}
\usepackage{bbm}
\usepackage{color}
\usepackage{cancel}
\usepackage{latexsym}
\usepackage{here} 
\usepackage{mathrsfs}
\usepackage{multirow}
\usepackage{lineno}
\usepackage{natbib}
\evensidemargin=\oddsidemargin

\newcommand{\tr}{\mbox{tr}}
\newcommand{\VEV}[1]{\langle #1 \rangle}

\begin{document}

\preprint{KEK-TH-1632}
\preprint{}

\title{
Phenomenology of Partially Composite Standard Model}

\author{Tomohiro Abe}
\affiliation{KEK Theory Center, Tsukuba, 305-0801, Japan}
\author{Ryuichiro Kitano}
\affiliation{KEK Theory Center, Tsukuba, 305-0801, Japan}
\affiliation{The Graduate
University for Advanced Studies (Sokendai), Tsukuba, 305-0801, Japan}

\begin{abstract}
We propose a model to describe the low energy physics of partially
 composite Standard Model, in which the electroweak sector in the
 Standard Model is weakly coupled to some strong dynamics.
The vector resonances in the strong sector are introduced as the
 effective degrees of freedom, $W'$ and $Z'$, which mixes with the $W$
 and $Z$ bosons through the electroweak symmetry breaking.
Through the coupling to the strong sector, the
Standard Model Higgs boson becomes partially composite, and its
 properties are modified.
We study the constraints from the electroweak precision data and
direct searches of $W'$ and $Z'$ at the LHC expriments, and discuss the
effects on the production/decay properties of the Higgs boson.
\end{abstract}

\maketitle
\section{Introduction}

The discovery of the Higgs boson at 125~GeV and the measurements of its
production and decay rates strongly support the picture of the weakly
coupled Higgs mechanism for the electroweak symmetry breaking.
It is then important to (re)consider the origin of the Higgs field and
its potential. 
The fact that the Higgs boson mass is much smaller than $4 \pi v$, where
$v=246$~GeV, 
tells us that, the dynamics behind electroweak symmetry
breaking has a very good description in terms of the linear sigma
model.

Not only from its mass, constraints from flavor changing neutral current
and $CP$ violating processes suggests that the picture of the elementary
Higgs field continues to be valid up to much higher than ${\cal
O}(10)$~TeV energy scale. This may be indicating that the physics to
reveal the origin of the Higgs field (which is possibly string theory)
is out of the reach of the LHC experiments.

Although the picture of the weakly coupled elementary Higgs field may be
valid up to a very high energy scale, it can be a different question
what generates the Higgs potential to drive the electroweak symmetry
breaking, and what sets the scale of the vacuum expectation value (VEV),
$v=246$~GeV.
For example, one can consider new particles or dynamics at a TeV energy
scale and the Higgs field couples to it so that the Higgs potential is
generated.
If the elementary Higgs field is weakly coupled to the TeV scale
dynamics, the Higgs field naturally obtains potential to explain
$v=246$~GeV, while the picture of the elementary Higgs fields remains
valid much above the TeV energy scale.
The coupling to the dynamical sector generically causes mixing between
the elementary Higgs and some (composite) operators in the dynamics,
making the observed Higgs boson a partially composite particle.
The picture that the Higgs field is weakly coupled to a TeV dynamics is
particularly motivated in supersymmetric models where the simplest
model, the MSSM, 
predicts the Higgs boson to be lighter than the $Z$ boson at tree
level. The partial compositeness explains why the Higgs boson is heavy
while the new dynamics can possibly provide a mechanism to address the
naturalness problem in supersymmetric models~\cite{Fukushima:2010pm,
Craig:2011ev, Csaki:2011xn, Azatov:2011ht, Azatov:2011ps,
Gherghetta:2011na, Heckman:2011bb, Csaki:2012fh, Evans:2012uf,
Kitano:2012wv, Kitano:2012cz}. See also Refs.~\cite{Samuel:1990dq,
Dine:1990jd, Harnik:2003rs} for earlier proposals of TeV scale
supersymmetric dynamics with elementary Higgs fields. 
For more ambitious proposals to break supersymmetry by the same
dynamics, see Refs.~\cite{Witten:1981nf, Dine:1981za, Dimopoulos:1981au}.
In these works, it
has been assumed that the dynamics rather than the Higgs VEV is the main
source for the electroweak symmetry breaking. Although such a
situation is now severely constrained by the electroweak precision tests,
the framework is still an attractive possibility by focusing on a different
region of the parameter space where the Higgs VEV is the main
contribution to the $W$ boson mass.
For proposals and studies of non-supersymmetric models,
see~\cite{Carone:1993xc, Simmons:1988fu, Carone:1993vg}.

In the TeV dynamics we consider, there should be Higgs-like operators
which can couple to the elementary Higgs field. This indicates that the
dynamical sector has SU(2)$_L \times$U(1)$_Y$ as a part of the global
symmetry just as in QCD. We, therefore, expect that there are
resonances, $W'$ and $Z'$, which couple to the SU(2)$_L \times$U(1)$_Y$
current, such as the $\rho$ meson in QCD.
Without specifying the actual dynamics, one can construct an effective
theory of vector resonances $W'$ and $Z'$ as the gauge bosons of
spontaneous broken gauge theory analogous to the Hidden Local
Symmetry~\cite{Bando:1987ym, Bando:1984ej, Bando:1984pw, Bando:1985rf}
in QCD.
In the partially composite Higgs framework, we expect that the vector
resonances appear at a TeV energy scale, which is within the reach of
the LHC experiments.

In this paper, we construct an effective theory of the $W'/Z'$ sector
which couples to the Standard Model Higgs boson.
The Higgs operators which give masses to $W'/Z'$ can mix with the
Standard Model Higgs boson, and triggers the electroweak symmetry
breaking.
We first examine the constraints from the electroweak precision tests,
and see which region of the parameter space is allowed.
We then reinterpret the results of the searches for $W'/Z'$ in the
sequential Standard Model at the LHC to the constraints on $W'/Z'$ in
the model. We see that the LHC experiments 
give stronger constraints than the precision tests in some parameter regions.
The properties of the Higgs boson are modified by the partial
compositeness. We examine whether such modification is allowed by the
present data.
For example, we find that there are
parameter regions where the strong sector components of the Higgs boson
is as large as $30~\%$.

Our work is closely related to Ref.~\cite{Bellazzini:2012tv} where
phenomenology of the models with $W'/Z'$ is studied motivated by the
framework of the Higgs boson as a pseudo Nambu-Goldstone
boson~\cite{ArkaniHamed:2001nc, Contino:2003ve}. There, a scalar
particle is added to the non-linear sigma models of $W'/Z'$ and discuss
the constraints from the electroweak precision measurements and from the
LHC data.
We, on the other hand, construct a linear sigma model to describe
$W'/Z'$ resonances and couple the Higgs field to it. Compared to the
work in Ref.~\cite{Bellazzini:2012tv}, 
we do not need to assume relations among parameters motivated by the
restoration of the perturbative unitarity, 
and 
all the physical quantities are, in
principle, calculable. In the linear sigma model, we find that there is
an important parameter, $r$, which describes the parity violation in the
dynamical sector. We see that the constraints from the precision
measurements prefer a large parity violation, and in such parameter
regions, the searches for $W'/Z'$ at the LHC experiments become more
important.

\section{Model}
\label{sec:model}

\subsection{Lagrangian}

We construct a model to describe the $W'$ and $Z'$ bosons as the gauge
bosons of new SU(2) gauge interactions.
The full gauge symmetry is, therefore,
SU(3)$\times$SU(2)$_{0}\times$SU(2)$_{1}\times$U(1)$_2$, where SU(3) is
QCD, and remaining parts are the electroweak sector.  
The SU(2)$_{1}$ gauge factor is the one which is analogous to the HLS in
QCD, and thus we assume its gauge coupling is much larger than those of
SU(2)$_{0}$ and U(1)$_2$.
We also assume that all the quarks and leptons are elementary. They do
not carry SU(2)$_1$ charges although they eventually couples to
$W'$/$Z'$ through mixing.
The left-handed fermions are fundamental representation of SU(2)$_{0}$,
and the right-handed fermions are singlet. 
All the fermions have appropriate U(1)$_2$ charges to reproduce the
electric charges.

\begin{figure}[tbp]
\begin{tabular}{c}
 \includegraphics[angle=0, width=0.4\hsize ]{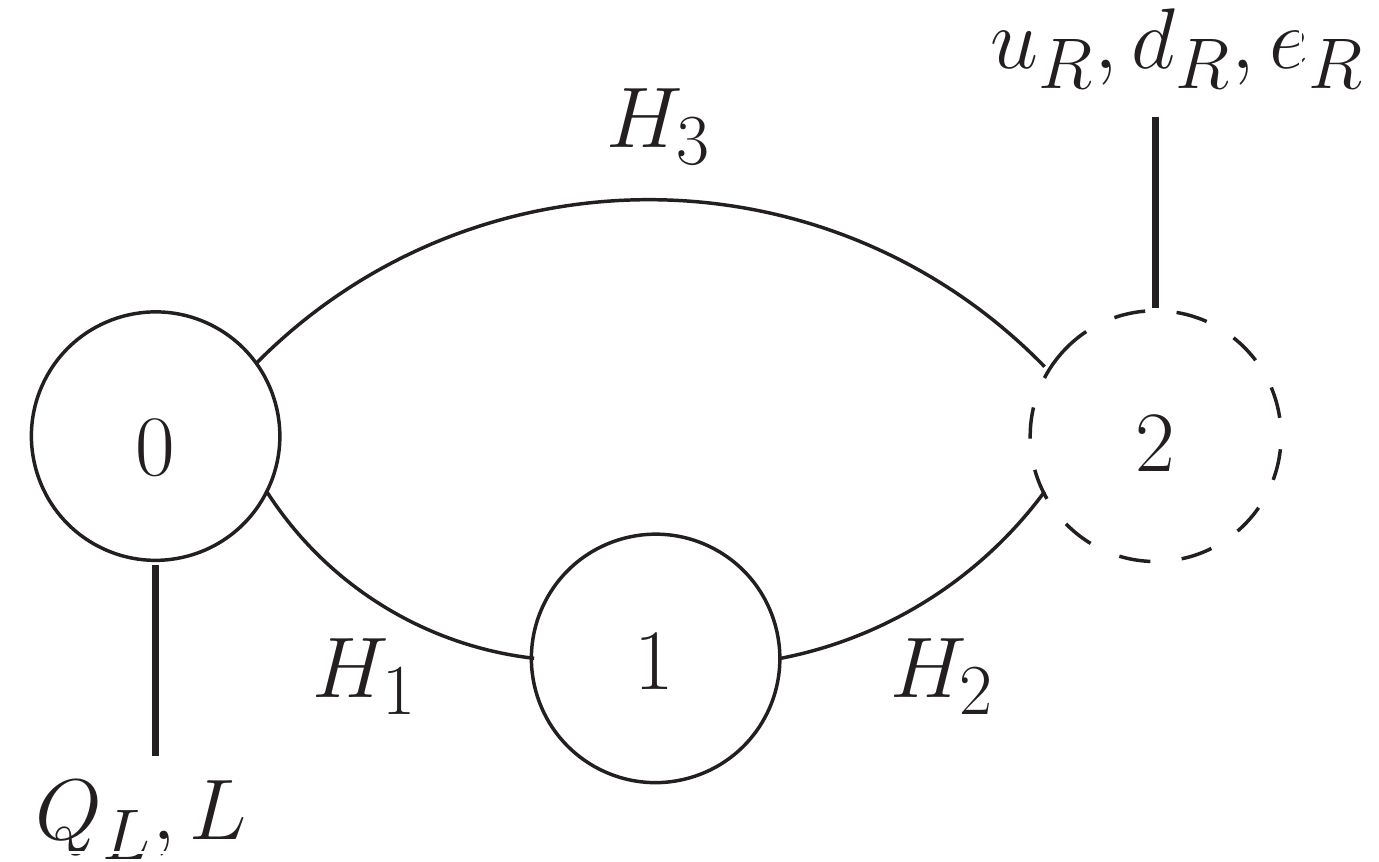}
\end{tabular}
 \caption{A schematic model description by the moose notation: The circles
 represent gauge symmetries. The dashed one is U(1) gauge symmetry. The
 lines connect two circles are the Higgs fields and break the symmetry
 they connects. The lines attached to the 0th and 2nd sites represent left-
 and right-handed fermions respectively. We assume that $H_1$, $H_2$,
 and the 1st site belong to the dynamical sector.
}
 \label{fig:moose}
 \end{figure}

Three Higgs fields, $H_1, H_2$, and $H_3$, are introduced for the
electroweak symmetry breaking.
The vacuum expectation values (VEVs) of $H_1, H_2$, and $H_3$ break
SU(2)$_0 \times$SU(2)$_1$, SU(2)$_1 \times$U(1)$_2$, and SU(2)$_0
\times$U(1)$_2$, respectively.\footnote{The same symmetry breaking
pattern is studied in Ref.~\cite{Chivukula:2009ck} in the top triangle
moose model.}
The $H_1$ and $H_2$ fields represent the condensations in the dynamical
sector. Their VEVs give masses to $W'$ and $Z'$. On the other hand,
$H_3$ is the elementary Higgs boson in the Standard Model. Through the
Higgs potential, the Standard Model Higgs field $H_3$ mixes with
``hadrons,'' ($H_1$ and $H_2$) in the dynamical sector, and thus becomes
partially composite.
All the fields except gauge bosons are summarized in
Table~\ref{tab:matters}.
We show a schematic description of the model in Fig.~\ref{fig:moose} by
using the moose notation~\cite{Georgi:1985hf}.
The model is simply the Standard Model added by $H_1$, $H_2$ and
the SU(2)$_1$ gauge fields.

Each Higgs fields contains four real scalars, and six of them are eaten
by the gauge bosons. So the six (= $4 \times 3$ $-$ 6) scalars remain as
physical degrees of freedom. This is the minimal model for the partially
composite Higgs boson.  Note that if we did not consider partial
compositeness with the same gauge symmetry, two Higgs fields would be
enough to break the symmetry~\cite{Hsieh:2010zr, Schmaltz:2010xr,
Grojean:2011vu, Abe:2012fb,Wang:2013jwa}.
In such models, physical charged scalar bosons are absent. However, in
our setup, there are charged and CP-odd scalars as well as CP-even
scalars. The existence of the charged and CP-odd scalar bosons are
distinctive feature of our model compared to other SU(2) models.

The models without $H_3$ are strongly constrained from the $S$/$T$
parameters. It has been observed that such constraints get significantly
weaker when the SM fermions are charged under
SU(2)$_1$, {\it i.e.}, where SM 
fermions are composite~\cite{Cacciapaglia:2004rb, Foadi:2004ps,
Chivukula:2005xm, Casalbuoni:2005rs, Chivukula:2005bn}. Such
models, if they exist, are subject to the 
constraints from searches for FCNC/CP non-conservations. In this paper,
we take a more conservative approach that the SM fermions are all
elementary and there is a fundamental Higgs field which give masses to
fermions through the Yukawa intereactions, so that the well-tested CKM
theory is not modified.
%
%
%
\begin{table}[tbp]
\begin{center}
\begin{tabular}{c|cccc}
fields & SU(2)$_0$ & SU(2)$_1$ & U(1)$_2$ & SU(3)$_c$\\
\hline
$H_1$ & $\bf 2$  & $\bf 2$ & 0 & 1  \\
$H_2$ & $\bf 1$  & $\bf 2$ & 1/2 & $\bf 1$  \\
$H_3$ & $\bf 2$  & $\bf 1$ & 1/2 & $\bf 1$  \\
\hline
$Q_{L}$ & $\bf 2$ & $\bf 1$ & 1/6 & $\bf 3$ \\
$L$ & $\bf 2$ & $\bf 1$ & -1/2 & $\bf 1$ \\
\hline
$u_R$ & $\bf 1$ & $\bf 1$ & 2/3 & $\bf 3$ \\
$d_R$ & $\bf 1$ & $\bf 1$ & -1/3 & $\bf 3$ \\
$e_R$ & $\bf 1$ & $\bf 1$ & -1 & $\bf 1$ \\
\end{tabular}
\caption{Quantum numbers of the Higgs and matter fields.}
\label{tab:matters}
\end{center}
\end{table}
%
%

The Lagrangian is given as follows:
\begin{align}
{\cal L}^{\text{gauge}}
= &
- \frac{1}{4} \sum_{a=1}^{3} W^{a}_{0 \mu \nu} W^{a \mu \nu}_{0}
- \frac{1}{4} \sum_{a=1}^{3} W^{a}_{1 \mu \nu} W^{a \mu \nu}_{1}
- \frac{1}{4} B_{\mu \nu} B^{\mu \nu}
,\\
{\cal L}^{\text{Higgs}}
= &
\tr\left( (D_{\mu} H_1)^{\dagger} D^{\mu} H_1\right)
+
\tr\left( (D_{\mu} H_2)^{\dagger} D^{\mu} H_2\right)
+
\tr\left( (D_{\mu} H_3)^{\dagger} D^{\mu} H_3\right)
-
V\left( H_1, H_2, H_3 \right)
,\\
{\cal L}^{\text{matter}}
= &
\sum_{i}
\left(
\overline{Q}_L^{i} i \gamma^{\mu} D_{\mu} Q_L^{i}
+
\overline{u}_R^{i} i \gamma^{\mu} D_{\mu} u_R^{i}
+
\overline{d}_R^{i} i \gamma^{\mu} D_{\mu} d_R^{i}
+
\overline{L}^{i} i \gamma^{\mu} D_{\mu} L^{i}
+
\overline{e}_R^{i} i \gamma^{\mu} D_{\mu} e_R^{i}
\right)
,\\
{\cal L}^{\text{Yukawa}}
= &
- 
\sum_{i,j}
\overline{Q}_L^{i} H_3
\left(
\begin{array}{cc}
 y_u^{ij}& 0 \\
 0 & y_d^{ij} 
\end{array}
\right)
\left(
\begin{array}{c}
 u_R^{j} \\
 d_R^{j}
\end{array}
\right)
-
\sum_{i}
 \overline{L}^{i} H_3
\left(
\begin{array}{cc}
 0& 0 \\
 0 & y_e^{i} 
\end{array}
\right)
\left(
\begin{array}{c}
 0\\
 e_R^{i}
\end{array}
\right)
+
(h.c.)
,
\end{align}
where
$i$ and $j$ are generation indices.
The Higgs fields are given by\footnote{$h_i$'s are proportional to 2 by 2
unit matrices though we do not write them explicitly.}
\begin{align}
 H_1
=&
\VEV{H_1}
+
\frac{1}{2}
\left(
h_1 + i \sum_{a=1}^{3}\tau^{a} \pi_1^a
\right)
, \\ 
 H_2
=&
\VEV{H_2}
+
\frac{1}{2}
\left(
h_2 + i \sum_{a=1}^{3}\tau^{a} \pi_2^a
\right)
, \\ 
 H_3
=&
\VEV{H_3}
+
\frac{1}{2}
\left(
h_3 + i \sum_{a=1}^{3}\tau^{a} \pi_3^a
\right)
,
\end{align}
where
$\tau^a$ denote the Pauli matrices, 
and $T^{a} = \tau^{a}/2$. Note that we take the matrix notation for the
Higgs fields. All the Higgs fields are under the constraint:
\begin{align}
 \tau^2 H_i^* \tau^2& = H_i, \quad i = 1,2,3.
\end{align}
The Higgs potential, $V(H_1, H_2, H_3)$, is
\begin{align}
 V(H_1, H_2, H_3)
=&
  \mu_1^2 \tr\left( H_1 H_1^{\dagger} \right)
+ \mu_2^2 \tr\left( H_2 H_2^{\dagger} \right)
+ \mu_3^2 \tr\left( H_3 H_3^{\dagger} \right)
\\
&
+
 \kappa \tr \left( H_1 H_2 H_3^{\dagger} \right)
\\
&
+ 
\lambda_1 
\left( \tr\left( H_1 H_1^{\dagger} \right) \right)^2
+ 
\lambda_2 
\left( \tr\left( H_2 H_2^{\dagger} \right) \right)^2
+ 
\lambda_3 
\left( \tr\left( H_3 H_3^{\dagger} \right) \right)^2
\\
&
+ 
\lambda_{12} 
\tr\left( H_1 H_1^{\dagger} \right)
\tr\left( H_2 H_2^{\dagger} \right)
+ 
\lambda_{23} 
\tr\left( H_2 H_2^{\dagger} \right)
\tr\left( H_3 H_3^{\dagger} \right)
+ 
\lambda_{31} 
\tr\left( H_3 H_3^{\dagger} \right)
\tr\left( H_1 H_1^{\dagger} \right)
.
\end{align}
Here all coefficients can be taken as real numbers.
Note that
\begin{align}
\left(
\tr \left( H_1 H_2 H_3^{\dagger} \right)
\right)^{*}
&=
\tr \left( H_1 H_2 H_3^{\dagger} \right)
.
\end{align}
We can also write the following term:
\begin{align}
\tr \left( H_1 H_2 \tau^3 H_3^{\dagger} \right)
.
\label{eq:kappa2}
\end{align}
This term can be eliminated by a field redefinition
of $H_2$.\footnote{A brief discussion is given in Appendix \ref{sec:kappa2}.}
Since the vacuum should respect U(1)$_{\text{em}}$ symmetry, the Higgs
VEVs, $\VEV{H_1}, \VEV{H_2}$, and $\VEV{H_3}$, should be diagonal.  In
addition, we can always take $\VEV{\pi_i^3}=0$ by the gauge
transformations.  So we work in a basis in which all the Higgs VEVs are
proportional to the identity matrix:
\begin{align}
& \VEV{H_1} = \frac{v_1}{2}
, \quad 
\VEV{H_2} = \frac{v_2}{2} 
, \quad 
\VEV{H_3} = \frac{v_3}{2} 
,
\label{eq:vacuum}
\end{align}
where $v_1, v_2,$ and $v_3$ are real and positive numbers.
We introduce $v$ and $r$ as
\begin{align}
 v^2
=&
 \frac{v_1^2 v_2^2}{v_1^2 + v_2^2}
+
 v_3^2
,
\quad
 r = \frac{v_2}{v_1}
\label{eq:v}
.
\end{align}
As we will discuss in Sec.~\ref{sec:ST}, the relation between $v$ and
the Fermi constant is the same as the one in the Standard Model, $v^2 =
(\sqrt{2} G_F)^{-1}$, so $v \sim 246$~GeV.
The parameter $1 - v_3^2 / v^2$ measures the size of the contribution to
the electroweak symmetry breaking from the dynamical sector. The ratio
$r$ is an important parameter in later discussion. In QCD-like
technicolor theories, $r=1$ is predicted due to the parity
conservation. As we see later, the model with $r=1$ is severely
constrained by the electroweak precision tests.

The limits $r = 0$ and $r \to \infty$ are other special points where
parity $(H_1 \leftrightarrow H_2)$ is maximally violating. Such points
can be the minimum of the potential when $\kappa = 0$, where an axial
U(1) symmetry, which is the one used to eliminate the term in
Eq.~\eqref{eq:kappa2}, is enhanced. For $r = 0$ or $r \to \infty$, which
means $v_2 = 0$ or $v_1 = 0$, the U(1) symmetry remains unbroken, and
thus there is no massless Nambu-Goldstone boson in the spectrum.
This vacuum realizes the Standard Model limit of the model, where there
is no mixing between $W/Z$ and $W'/Z'$, and $H_3$ is the only source of
the electroweak symmetry breaking $(v_3 = v)$.
When a small $\kappa$ parameter is turned on, one can naturally realize
$r \ll 1$ or $r \gg 1$. Since such parameter regions are close to the
Standard Model limit, the constraints from the electroweak precision
tests are not very severe. However, as we will see later, the searches
for $W'/Z'$ at the LHC experiments become important in such parameter
regions.

The parameter $r$ is related to the custodial symmetry: the symmetry
between $W (W')$ and $Z (Z')$.
The custodial symmetry becomes a good symmetry for a small $r$.  This
can be understood by the nature of $W'$ and $Z'$. For $r \ll 1$, both
$W'$ and $Z'$ mainly originate from SU(2)$_1$, whereas for $r \gg 1$,
$Z'$ has a large U(1)$_2$ fraction which $W'$ does not have.
We will explicitly see this feature, for example, in
Sec.~\ref{sec:directDetection}.

In general, demanding the vacuum in Eq.~\eqref{eq:vacuum} as a extremum
of the potential, we obtain the following relations:
\begin{align}
 \mu_1^2
&=
-
\kappa
\frac{v_2 v_3}{4 v_1}
-
\frac{1}{2}
\left(
2 v_1^2 \lambda_1
+ v_2^2 \lambda_{12}
+ v_3^2 \lambda_{31}
\right)
, \\ 
 \mu_2^2
&=
-
\kappa
\frac{v_3 v_1}{v_2}
-
\frac{1}{2}
\left(
v_1^2 \lambda_{12}
+
2 v_2^2 \lambda_2 
+
v_3^2 \lambda_{23}
\right)
, \\ 
 \mu_3^2
&=
-
\kappa
\frac{v_1 v_2}{4 v_3}
-
\frac{1}{2}
\left(
v_1^2 \lambda_{31}
+
v_2^2 \lambda_{23}
+
2 v_3^2 \lambda_3
\right)
.
\end{align}
For the stability of the potential at a large value of the Higgs fields,
the following relations should be satisfied:
\begin{align}
&
 \lambda_1 >0, \quad
 \lambda_2 >0, \quad
 \lambda_3 >0.
\end{align}
We will also see that 
\begin{align}
& \kappa < 0
\end{align}
is required from the local stability when $v_1, v_2, v_3 \neq 0$.

\subsection{Higgs mass}
From the Higgs potential, we can read off the following mass terms for
the physical scalar particles.

\subsubsection{charged Higgs sector}
The mass matrix of the charged Higgs fields is given by
\begin{align}
V
\supset
\left(
\begin{matrix}
 \pi_1^{+} & \pi_2^{+} & \pi_3^{+}
\end{matrix}
\right)
{\cal M}_{\text{CS}}^2
\left(
\begin{matrix}
 \pi_1^{-} \\ \pi_2^{-} \\ \pi_3^{-}
\end{matrix}
\right)
=
\left(
\begin{matrix}
 \pi_{W^{+}} & \pi_{{W}^{\prime +}} & H^{+}
\end{matrix}
\right)
\left(
\begin{matrix}
 0 & 0 & 0 \\
 0 & 0 & 0 \\
 0 & 0 & m_{H^{\pm}}^2 
\end{matrix}
\right)
\left(
\begin{matrix}
 \pi_{{W}^{-}} \\ \pi_{W^{\prime -}} \\ H^{-}
\end{matrix}
\right)
,
\end{align}
where
\begin{align}
{\cal M}_{\text{CS}}^2
=&
\left(
\begin{matrix}
  -\frac{v_2 v_3}{2 v_1} \kappa
& -\frac{v_3}{2} \kappa
& \frac{v_2}{2} \kappa 
\\
  -\frac{v_3}{2} \kappa
& -\frac{v_1 v_3}{2 v_2} \kappa
& \frac{v_1}{2} \kappa
\\
  \frac{v_2}{2} \kappa
& \frac{v_1}{2} \kappa
& -\frac{v_1 v_2}{2 v_3} \kappa
\end{matrix}
\right)
, \\ 
 m_{H^{\pm}}^2
=&
-2 \frac{\kappa}{v_3}
\frac{1+r^2}{r}
v^2
.
\label{eq:chargedHiggsMass}
\end{align}
The fields $\pi_{{W}^{\pm}}$ and $\pi_{{W'}^{\pm}}$ are would-be NG
bosons which are eaten by $W$ and $W'$ respectively.
The relation between mass eigenstates and gauge eigenstates are
\begin{align}
 H^{\pm}
=&
 w_{A}^{1} \pi_1^{\pm}
+
 w_{A}^{2} \pi_2^{\pm}
+
 w_{A}^{3} \pi_3^{\pm}
,
\end{align}
where
\begin{align}
 w_{A}^{1}
=&
\frac{1}{\sqrt{\frac{1}{v_1^2} + \frac{1}{v_2^2} + \frac{1}{v_3^2}}}
 \frac{1}{v_1}
=
\frac{r}{\sqrt{1+r^2}} 
\frac{v_3}{v}
\label{eq:vH1}
, \\ 
 w_{A}^{2}
=&
\frac{1}{\sqrt{\frac{1}{v_1^2} + \frac{1}{v_2^2} + \frac{1}{v_3^2}}}
 \frac{1}{v_2}
=
\frac{1}{\sqrt{1+r^2}} 
\frac{v_3}{v}
, \\ 
 w_{A}^{3}
=&
-
\frac{1}{\sqrt{\frac{1}{v_1^2} + \frac{1}{v_2^2} + \frac{1}{v_3^2}}}
 \frac{1}{v_3}
=
-
\sqrt{1-\frac{v_3^2}{v^2}}
.
\label{eq:vH3}
\end{align}

\subsubsection{neutral CP odd Higgs sector}
The mass matrix for the CP-odd scalar fields is given by
\begin{align}
V
\supset
\frac{1}{2}
\left(
\begin{matrix}
 \pi_1^{3} & \pi_2^{3} & \pi_3^{3}
\end{matrix}
\right)
{\cal M}_{\text{NS}}^2
\left(
\begin{matrix}
 \pi_1^{3} \\ \pi_2^{3} \\ \pi_3^{3}
\end{matrix}
\right)
=
\left(
\begin{matrix}
 \pi_{Z} & \pi_{Z'} & A^{0}
\end{matrix}
\right)
\left(
\begin{matrix}
 0 & 0 & 0 \\
 0 & 0 & 0 \\
 0 & 0 & m_{A^{0}}^2 
\end{matrix}
\right)
\left(
\begin{matrix}
 \pi_{Z} \\ \pi_{Z'} \\ A^{0}
\end{matrix}
\right)
,
\end{align}
where
\begin{align}
{\cal M}_{\text{NS}}^2
=&
{\cal M}_{\text{CS}}^2
, \\ 
 m_{A^{0}}^2
=&
 m_{H^{\pm}}^2
.
\end{align}
The physical CP-odd Higgs boson has the same mass as the charged Higgs
boson given in Eq.~(\ref{eq:chargedHiggsMass}). The fields $\pi_{Z}$ and
$\pi_{Z'}$ are would-be NG bosons which are eaten by $Z$ and $Z'$
respectively.
The relation between mass eigenstates and gauge eigenstates are
\begin{align}
 A^{0}
=&
 w_{A}^{1} \pi_1^{3}
+
 w_{A}^{2} \pi_2^{3}
+
 w_{A}^{3} \pi_3^{3}
,
\end{align}
where
$w_{A}^{1}, w_{A}^{2}$, and $w_{A}^{3}$ are given in
Eqs.~(\ref{eq:vH1})--(\ref{eq:vH3}).

\subsubsection{neutral CP even Higgs sector}

The mass matrix for the neutral CP even Higgs bosons is
\begin{align}
V
\supset
\frac{1}{2}
\left(
\begin{matrix}
 h_1 & h_2 & h_3
\end{matrix}
\right)
{\cal M}_{\text{H}}^2
\left(
\begin{matrix}
 h_1 \\ h_2 \\ h_3
\end{matrix}
\right)
=
\left(
\begin{matrix}
 H & H' & h 
\end{matrix}
\right)
\left(
\begin{matrix}
 m_H^2 & 0 & 0 \\
 0 & m_{H'}^2 & 0 \\
 0 & 0 & m_h^2 
\end{matrix}
\right)
\left(
\begin{matrix}
 H \\ H' \\ h
\end{matrix}
\right)
,
\end{align}
where
\begin{align}
{\cal M}_{\text{H}}^2
= &
\left(
\begin{matrix}
 -\frac{v_2 v_3}{2 v_1} \kappa  + 4 v_1^2 \lambda_1 &
 \frac{v_3}{2} \kappa + 2 v_1 v_2 \lambda_{12} &
 \frac{v_2}{2} \kappa + 2 v_1 v_3 \lambda_{31} \\
 \frac{v_3}{2} \kappa + 2 v_1 v_2 \lambda_{12} &
 -\frac{v_1 v_3}{2 v_2} \kappa  + 4 v_2^2 \lambda_2 &
 \frac{v_1}{2} \kappa + 2 v_2 v_3 \lambda_{23} \\
 \frac{v_1}{2} \kappa + 2 v_2 v_3 \lambda_{23} &
 \frac{v_1}{2} \kappa + 2 v_2 v_3 \lambda_{23} &
 -\frac{v_1 v_2}{2 v_3} \kappa  + 4 v_3^2 \lambda_3 
\end{matrix}
\right)
.
\end{align}
The relation between mass eigenstates and gauge eigenstates are
\begin{align}
\left(
\begin{matrix}
 H & H' & h
\end{matrix}
\right)
=&
\left(
\begin{matrix}
 w_H^{1} & w_H^{2} & w_H^{3} \\
 w_{H'}^{1} & w_{H'}^{2} & w_{H'}^{3} \\
 w_h^{1} & w_h^{2} & w_h^{3} \\
\end{matrix}
\right)
\left(
\begin{matrix}
 h_1 \\ h_2 \\ h_3
\end{matrix}
\right)
,
\quad 
\left(
\begin{matrix}
 h_1 & h_2 & h_3
\end{matrix}
\right)
=
\left(
\begin{matrix}
 w_H^{1} & w_{H'}^{1} & w_h^{1} \\
 w_H^{2} & w_{H'}^{2} & w_h^{2} \\
 w_H^{3} & w_{H'}^{3} & w_h^{3} \\
\end{matrix}
\right)
\left(
\begin{matrix}
 H \\ H' \\ h
\end{matrix}
\right)
.
\end{align}
We define $h$ as the lightest one and thus $m_h = $125~GeV.  Note that
$|w_{h}^{i}|^2$ is the $h$ component in $H_i$. Since we assume $H_3$ is
elementary and $H_1$ and $H_2$ are composite, $|w_{h}^3|=1$ means
$h$ is completely elementary. If $h$ is completely composite and arise
from the dynamical sector, then $|w_{h}^3|=0$. Our focus is partially
composite $h$, {\it i.e.},
\begin{align}
 |w_{h}^{3}| \neq 1,
\quad
 |w_{h}^{3}|^2 \gg |w_{h}^{1}|^2, 
\quad 
 |w_{h}^{3}|^2 \gg|w_{h}^{2}|^2.
\label{eq:partially_compositeness}
\end{align}

\subsection{Gauge sector}

The mass terms of the gauge bosons are
\begin{align}
 {\cal L}
\supset&
\left(
\begin{matrix}
 W_{0 \mu}^{+} & W_{1 \mu}^{+}
\end{matrix}
\right)
{\cal M}_{\text{CG}}^2
\left(
\begin{matrix}
 W_0^{- \mu} \\
 W_1^{- \mu}
\end{matrix}
\right)
+
\frac{1}{2}
\left(
\begin{matrix}
 W_{0 \mu}^{3} & W_{1 \mu}^{3} & B_{\mu}
\end{matrix}
\right)
{\cal M}_{\text{NG}}^2
\left(
\begin{matrix}
 W_0^{3 \mu} \\
 W_1^{3 \mu} \\
 B^{3}
\end{matrix}
\right)
,
\end{align}
where
\begin{align}
{\cal M}_{\text{CG}}^2
= &
\frac{1}{4}
\left(
\begin{matrix}
 g_0^2 (v_1^2 + v_3^2) & -g_0 g_1 v_1^2 \\
 -g_0 g_1 v_1^2 & g_1^2 ( v_2^2 + v_3^2) 
\end{matrix}
\right)
, \\ 
{\cal M}_{\text{NG}}^2
= &
\frac{1}{4}
\left(
\begin{matrix}
 g_0^2 (v_1^2 + v_3^2) & -g_0 g_1 v_1^2 & -g_0 g_2 v_3^2 \\
 -g_0 g_1 v_1^2 & g_1^2 ( v_2^2 + v_3^2) & -g_1 g_2 v_3^2 \\
 -g_0 g_2 v_3^2 & -g_1 g_2 v_3^2 & g_2^2 (v_2^2 + v_3^2)
\end{matrix}
\right)
.
\end{align}
The gauge boson masses are the eigenvalues of these mass matrices. In
 the $g_1 \gg g_{0, 2}$ region, we find
\begin{align}
 m_{W}^2
\simeq&
 \frac{1}{4}
 g_0^2 v^2
\left(
1
-
  \frac{g_0^2}{g_1^2} \frac{1}{(1 + r^2)^2}
\right)
, \\ 
 m_{W'}^2
\simeq&
 \frac{1}{4}
 g_1^2 (v_1^2 + v_2^2)
\left(
1
+
 \frac{g_0^2}{g_1^2} \frac{1}{(1 + r^2)^2}
\right)
\label{eq:W'mass}
, \\ 
m_{\gamma}^2
=&
0
, \\ 
 m_{Z}^2
\simeq&
 \frac{1}{4}
 (g_0^2 + g_2^2) v^2
\left(
1
-
\frac{(g_0^2 - g_2^2 r^2)^2}{g_1^2 (g_0^2 + g_2^2)}
\frac{1}{(1 + r^2)^2}
\right)
, \\ 
 m_{Z'}^2
\simeq&
 \frac{1}{4}
 g_1^2 (v_1^2 + v_2^2)
\left(
1
+
 \frac{g_0^2 + g_2^2 r^4}{g_1^2} \frac{1}{(1 + r^2)^2}
\right)
\label{eq:Z'mass}
.
\end{align}
The relation between mass eigenstates and gauge eigenstates are
\begin{align}
\left(
 \begin{matrix}
  W_{\mu}^{\pm} \\
  W_{\mu}^{\prime \pm}
 \end{matrix}
\right)
=
\left(
\begin{matrix}
 w_W^{0} & w_W^{1} \\
 w_{W'}^{0} & w_{W'}^{1}
\end{matrix}
\right)
\left(
 \begin{matrix}
  W_{0 \mu}^{\pm} \\
  W_{1 \mu}^{\pm}
 \end{matrix}
\right)
,
\quad
&
\left(
 \begin{matrix}
  W_{0 \mu}^{\pm} \\
  W_{1 \mu}^{\pm}
 \end{matrix}
\right)
=
\left(
\begin{matrix}
 w_W^{0} & w_{W'}^{0} \\
 w_{W}^{1} & w_{W'}^{1}
\end{matrix}
\right)
\left(
 \begin{matrix}
  W_{\mu}^{\pm} \\
  W_{\mu}^{\prime \pm}
 \end{matrix}
\right)
,
\\ 
\left(
 \begin{matrix}
  Z_{\mu} \\
  Z_{\mu}^{\prime} \\
  A_{\mu}
 \end{matrix}
\right)
=
\left(
\begin{matrix}
 w_Z^{0} & w_Z^{1} & w_Z^{2}\\
 w_{Z'}^{0} & w_{Z'}^{1} & w_{Z'}^{2} \\
 w_{A}^{0} & w_{A}^{1} & w_{A}^{2} \\
\end{matrix}
\right)
\left(
 \begin{matrix}
  W_{0 \mu}^{3} \\
  W_{1 \mu}^{3} \\
  B_{\mu} \\
 \end{matrix}
\right)
, 
\quad
&
\left(
 \begin{matrix}
  W_{0 \mu}^{3} \\
  W_{1 \mu}^{3} \\
  B_{\mu} \\
 \end{matrix}
\right)
=
\left(
\begin{matrix}
 w_Z^{0} & w_{Z'}^{0} & w_A^{0}\\
 w_Z^{1} & w_{Z'}^{1} & w_A^{1}\\
 w_Z^{2} & w_{Z'}^{2} & w_A^{2}
\end{matrix}
\right)
\left(
 \begin{matrix}
  Z_{\mu} \\
  Z_{\mu}^{\prime} \\
  A_{\mu}
 \end{matrix}
\right)
.
\end{align}
We can find the expressions of $w_{X}^{i}$ by diagonalizing the mass
matrices.

We find some relations among parameters.
A naive relation among $m_{W}$ and $m_{W'}$ is $m_{W'}^2 - m_{W}^2
>0$. But there is actually a more stringent bound:
\begin{align}
m_{W'}^2 - m_{W}^2
\geq&
\frac{2 m_W m_{W'}}{r}
\sqrt{1 - \frac{v_3^2}{v^2}}
.
\label{eq:lower_r}
\end{align}
We can use this relation to find the lower bound on $r$,
\begin{align}
r
\geq&
\frac{2 m_W m_{W'}}{m_{W'}^2 - m_{W}^2}
\sqrt{1 - \frac{v_3^2}{v^2}}
.
\end{align}
We find another relation,
\begin{align}
 g_1
<&
\frac{m_{W'}}{\sqrt{v^2 - v_3^2}}
.
\label{eq:upper_g1}
\end{align}
We derive Eqs.~(\ref{eq:lower_r}) and (\ref{eq:upper_g1}) in
Appendix~\ref{sec:gaugeDetails}.

\subsection{couplings}
\label{sec:couplings}
In this section, we calculate the coupling constants between mass
eigenstates. Since many of their expressions are complicated, we
use the approximation which are valid when $g_1 \gg g_0$.
We do not use this approximation in the numerical calculations performed
later.

\subsubsection{$h$-$f$-$f$ couplings}
The $h$ couplings to the fermions, and the fermion masses are given as
\begin{align}
 g_{ffh}
=&
 y_f \frac{w_h^{3}}{2}
, \quad
 m_{f}
=
 y_f \frac{v_3}{2}
.
\end{align}
From these formulae, we see that Yukawa couplings are always
$(v/v_3)$ times as large as their SM values. Since $v_3 < v \sim
246$~GeV as we can see from Eq.~(\ref{eq:v}), the Yukawa coupling
constants are always larger than their SM values. In order for the top
Yukawa coupling to be small enough for perturbative calculations, too
small $v_3$ is not allowed. For example, if we impose $y_t < 3
y_t^{\text{SM}}$ then the lower bound on $v_3$ is $\sim 80$~GeV.

We introduce $\kappa_f$ as the ratio of this coupling to the one in the
SM,
\begin{align}
\kappa_f
\equiv
 \frac{g_{ffh}}{m_f/v}
=&
 \frac{v}{v_3} w_h^{3}
.
\end{align}
Since $v_3 < v$, or the Yukawa couplings are larger than their SM
values, this ratio is larger than one when $w_h^3 > v_3 / v$. This
leads the enhancement of Br$(h \to ff)$.  We do not consider extremely
small values of $w_h^{3}$ (see Eq.~(\ref{eq:partially_compositeness})).
The choice is phenomenologically favored since the signal strengths around
125~GeV in both ATLAS and CMS look consistent with the SM Higgs boson.

\subsubsection{$V$-$f$-$f$ couplings}

The gauge boson to fermion couplings are given by
\begin{align}
 g_{Zff}
\simeq&
\frac{e}{s_Z c_Z}
\left(
T^3
-
s_Z^2
\left(
1
+
\frac{m_W^2}{m_{W'}^2}
\frac{1}{1 - 2 s_Z^2}
\left(
1 - \frac{v_3^2}{v^2}
\right)
\right)
Q
\right)
, \\ 
 g_{Z'ff}
\simeq&
-
\frac{e}{s_Z}
\frac{m_W}{m_{W'}}
\frac{1}{r}
\sqrt{1 - \frac{v_3^2}{v^2}}
\left(
\left(
1
-
r^2 \frac{s_Z^2}{c_Z^2}
\right)
T^3
+
r^2
\frac{s_Z^2}{c_Z^2}
Q
\right)
, \\ 
g_{Wff}
\simeq&
\frac{e}{s_Z}
\left(
1
-
\frac{m_W^2}{m_{W'}^2}
\frac{s_Z^2}{1 - 2 s_Z^2}
\left(
1 - \frac{v_3^2}{v^2}
\right)
\right)
, \\ 
g_{W'ff}
\simeq&
-
\frac{e}{s_Z}
\frac{m_W}{m_{W'}}
\frac{1}{r}
\sqrt{1 - \frac{v_3^2}{v^2}}
,
\end{align}
where
\begin{align}
 \frac{1}{e^2}
\equiv&
 \frac{1}{g_0^2}
+
 \frac{1}{g_1^2}
+
 \frac{1}{g_2^2}
,
\end{align}
and where $s_Z$ and $c_Z$ is defined through
\begin{align}
 s_Z^2 c_Z^2
\equiv&
 \frac{e^2}{4\sqrt{2} G_F m_Z^2} 
.
\end{align}

\subsubsection{$h$-$V$-$V$ couplings}
The Higgs boson couplings to the gauge bosons are given as follows.
\begin{align}
\kappa_W
\equiv
 \frac{g_{WWh}}{2 m_W^2 /v }
\simeq&
+
\frac{r}{(1+r^2)^{3/2}}
\sqrt{1 - \frac{v_3^2}{v^2}}
\left(
r^2
-
2 \frac{m_W^2}{m_{W'}^2}
\right)
w_h^{1}
\nonumber \\ 
&+
\frac{1}{(1+r^2)^{3/2}}
\sqrt{1 - \frac{v_3^2}{v^2}}
\left(
1
+
2 \frac{m_W^2}{m_{W'}^2}
\right)
w_h^{2}
\nonumber \\ 
&+
\frac{v_3}{v}
w_h^{3}
, \\ 
\kappa_{W'}
\equiv
 \frac{g_{W'W'h}}{2 m_{W'}^2 /v }
\simeq&
+
\frac{r}{(1+r^2)^{3/2}}
\frac{1}{\sqrt{1-\frac{v_3^2}{v^2}}}
\left(
1
+
2 
\frac{m_W^2}{m_{W'}^2}
\left(
1
-
\frac{v_3^2}{v^2}
\right)
\right)
w_h^{1}
\nonumber \\ 
&
+
\frac{1}{(1+r^2)^{3/2}}
\frac{1}{\sqrt{1-\frac{v_3^2}{v^2}}}
\left(
r^2
-
2 
\frac{m_W^2}{m_{W'}^2}
\left(
1
-
\frac{v_3^2}{v^2}
\right)
\right)
w_h^{2}
.
\end{align}
Here we ignored ${\cal O}\left( m_W^4/ m_{W'}^4 \right)$ terms.
Note that $h_3$ component in $h$, $w_h^3$, does not contribute to $g_{W'W'h}$,
namely $h_3$ does not couple to $W'W'$, at this order. 

In order to see their qualitative feature, let us consider the case with
$w_h^{1} \sim w_h^{2} \sim 0$ and $w_h^{3} \sim 1$, namely the situation
where the Higgs boson is almost elementary. In this case, we see that
$\kappa_W \sim v_3/v \leq 1$, and Br$(h \to WW)$ tends to be smaller
than the SM prediction. In addition, we find $\kappa_{W'} \sim 0$ in that
case, so the $W'$-loop effect on the $h \to \gamma \gamma $ process
tends to be small due to the small fractions of $w_h^{1}$ and $w_{h}^2$.
We can introduce $\kappa_{Z}$ and $\kappa_{Z'}$ in a similar manner.
\begin{align}
\kappa_{Z}
\equiv
 \frac{g_{ZZh}}{2 m_Z^2 /v }
\simeq&
+
\frac{r}{(1+r^2)^{3/2}}
\sqrt{1 - \frac{v_3^2}{v^2}}
\left(
r^2
-
2 \frac{m_W^2}{m_{W'}^2}
\left(
1
-
r^2 \frac{s_Z^2}{c_Z^2}
\right)
\right)
w_h^{1}
\nonumber \\ 
&+
\frac{1}{(1+r^2)^{3/2}}
\sqrt{1 - \frac{v_3^2}{v^2}}
\left(
1
+
2 \frac{m_W^2}{m_{W'}^2}
\left(
1
-
r^2 \frac{s_Z^2}{c_Z^2}
\right)
\right)
w_h^{2}
\nonumber \\ 
&+
\frac{v_3}{v}
w_h^{3}
, \\ 
\kappa_{Z'}
\equiv
 \frac{g_{Z'Z'h}}{2 m_{Z'}^2 /v }
\simeq&
+
\frac{r}{(1+r^2)^{3/2}}
\frac{1}{\sqrt{1-\frac{v_3^2}{v^2}}}
\left(
1
+
2 
\frac{m_W^2}{m_{W'}^2}
\left(
1
-
\frac{v_3^2}{v^2}
\right)
\left(
1
-
r^2 \frac{s_Z^2}{c_Z^2}
\right)
\right)
w_h^{1}
\nonumber \\ 
&
+
\frac{1}{(1+r^2)^{3/2}}
\frac{1}{\sqrt{1-\frac{v_3^2}{v^2}}}
\left(
r^2
-
2 
\frac{m_W^2}{m_{W'}^2}
\left(
1
-
\frac{v_3^2}{v^2}
\right)
\left(
1
-
r^2 \frac{s_Z^2}{c_Z^2}
\right)
\right)
w_h^{2}
.
\end{align}
The $g_{VV'h}$ couplings are also calculated to be:
\begin{align}
 \frac{g_{WW'h}}{2 m_{W} m_{W'} /v }
\simeq&
-
\frac{1}{(1+r^2)^{3/2}}
\left(
r^2
-
\frac{m_W^2}{m_{W'}^2}
\left(
(1-r^2) + r^2 \frac{v_3^2}{v^2}
\right)
\right)
w_h^{1}
\nonumber \\ 
&+
\frac{1}{r (1+r^2)^{3/2}}
\left(
r^2
-
\frac{m_W^2}{m_{W'}^2}
\left(
(1-r^2) - \frac{v_3^2}{v^2}
\right)
\right)
w_h^{2}
\nonumber \\ 
&-
\frac{m_W^2}{m_{W'}^2}
\frac{v_3}{v}
\frac{1}{r}
\sqrt{1-\frac{v_3^2}{v^2}}
w_h^{3}
, \\ 
 \frac{g_{ZZ'h}}{2 m_{Z} m_{Z'} /v }
\simeq&
-
\frac{1}{(1+r^2)^{3/2}}
\left(
r^2
-
\frac{m_W^2}{m_{W'}^2}
\left(
(1-r^2) + r^2 \frac{v_3^2}{v^2}
\right)
\left(
1
-
r^2 \frac{s_Z^2}{c_Z^2}
\right)
\right)
w_h^{1}
\nonumber \\ 
&+
\frac{1}{r (1+r^2)^{3/2}}
\left(
r^2
-
\frac{m_W^2}{m_{W'}^2}
\left(
(1-r^2) - \frac{v_3^2}{v^2}
\right)
\left(
1
-
r^2 \frac{s_Z^2}{c_Z^2}
\right)
\right)
w_h^{2}
\nonumber \\ 
&-
\frac{m_W^2}{m_{W'}^2}
\frac{v_3}{v}
\frac{1}{r}
\sqrt{1-\frac{v_3^2}{v^2}}
\left(
1
-
r^2 \frac{s_Z^2}{c_Z^2}
\right)
w_h^{3}
.
\end{align}
From these expressions, we see that the difference between $W (W')$ and
$Z (Z')$ becomes larger (smaller) when $r > 1 ~(r < 1)$.

\subsubsection{$h$-$H^{-}$-$H^{+}$ couplings}
The coupling between the Higgs boson and the charged Higgs bosons are 
\begin{align}
 {\cal L}
\supset&
 - g_{H^{-}H^{+}h} H^{+} H^{-} h
,
\end{align}
where
\begin{align}
 g_{H^{-}H^{+}h}
\simeq &
+
\frac{m_{H^{\pm}}^2}{v}
\frac{v_3^2}{v^2}
\left(
\sqrt{1 - \frac{v_3^2}{v^2}}
\frac{r}{(1+r^2)^{3/2}}
w_h^1
+
\sqrt{1 - \frac{v_3^2}{v^2}}
\frac{1}{(1+r^2)^{3/2}}
w_h^2
+
\frac{v_3}{v}
\frac{r^2}{(1 + r^2)^2}
w_h^3
\right)
.
\end{align}
Here we assume $m_{H^{\pm}}^2 \gg v^2, v_3^2 $.
We define $\kappa_{H^{\pm}}$ as follows;
\begin{align}
 \kappa_{H^{\pm}}
\equiv&
\frac{g_{H^{-}H^{+}h}}{2 m_{H^{\pm}}^2/v}
.
\end{align}

\subsubsection{$WWZ$, $WW'Z$, and $WWZ'$couplings}
Finally, the triple gauge boson vertices are given by
\begin{align}
 g_{WWZ}
\simeq&
\frac{e}{s_Z} c_Z
\left(
1
-
\frac{m_W^2}{m_{W'}^2}
\frac{1}{(1+r^2) ( 1 - 2 s_Z^2)}
\left(
1
+
r^2 \frac{s_Z^2}{c_Z^2}
\right)
\left(
1 - \frac{v_3^2}{v^2}
\right)
\right)
, \\ 
 g_{WW'Z}
\simeq&
-
\frac{e}{s_Z} c_Z
\frac{m_W}{m_{W'}}
\frac{r}{(1+r^2) (1 - s_Z^2)}
\sqrt{1 - \frac{v_3^2}{v^2}}
, \\ 
 g_{WWZ'}
\simeq&
-
\frac{e}{s_Z} c_Z
\frac{m_W}{m_{W'}}
\frac{r}{(1+r^2) (1 - s_Z^2)}
\sqrt{1 - \frac{v_3^2}{v^2}}
\sqrt{1-s_Z^2}
.
\end{align}
We find the $VVV'$ couplings are suppressed by $(m_W/m_{W'})$ compared to the
$WWZ$ coupling.

\section{Constraints on $W'$ and $Z'$}
\subsection{Constraints from electroweak precision measurements}
\label{sec:ST}
In this section, we discuss the electroweak constraints.
Due to the existence of the extra gauge bosons, $W'$ and $Z'$, the
parameters such as $\hat{S}$ are non-zero at tree level, and gives a
severe constraint on the model because these parameters are measured to
be at most as small as of order the one-loop level.

To calculate the electroweak parameters, $\hat{S}$, $\hat T$, $\hat U$,
$W$, and $Y$ (see Ref.~~\cite{Barbieri:2004qk} for definitions), we
calculate the quadratic terms of gauge bosons in the momentum-space
effective action.  They are written as
\begin{align}
-
\frac{1}{2}
g_{\mu \nu}
\left(
\begin{array}{ccc}
W_{0}^{\mu} & B^{\mu} & W_{1}^{\mu} \\
\end{array}
\right)
\left(
\begin{array}{ccc}
 \Pi_{00} & \Pi_{02} & \Pi_{01}\\
 \Pi_{20} & \Pi_{22} & \Pi_{21}\\
 \Pi_{10} & \Pi_{12} & \Pi_{11}\\
\end{array}
\right)
\left(
\begin{array}{c}
 W_{0}^{\nu}\\
 B^{\nu} \\
 W_{1}^{\nu}
\end{array}
\right)
+
(q_{\mu} q_{\nu} \text{ terms}),
\end{align}
where
\begin{align}
 \Pi_{00}
=&
 \frac{1}{g_0^2} q^2
-
 \frac{1}{4} (v_1^2 + v_3^2)
, \\ 
 \Pi_{22}
=&
 \frac{1}{g_2^2} q^2
-
 \frac{1}{4} (v_2^2 + v_3^2)
, \\ 
 \Pi_{11}
=&
 \frac{1}{g_1^2} q^2
-
 \frac{1}{4} (v_1^2 + v_2^2)
, \\ 
 \Pi_{01}
=&
 \frac{1}{4} v_1^2
, \\ 
 \Pi_{02}
=&
 \frac{1}{4} v_3^2
, \\ 
 \Pi_{12}
=&
 \frac{1}{4} v_2^2
,
\end{align}
and $\Pi_{xy} = \Pi_{yx}$.
We consider only $g_{\mu \nu}$ terms.
Since $W_{1}$ is decoupled from the fermion sector, we integrate it
out. Then the quadratic terms become
\begin{align}
 &
-
\frac{1}{2}
g_{\mu \nu}
\left(
\begin{array}{cc}
W_{0}^{\mu} & B^{\mu} 
\end{array}
\right)
\left(
\begin{array}{cc}
 \Pi_{W_3 W_3 } & \Pi_{W_3 B} \\
 \Pi_{W_3 B} & \Pi_{BB} \\
\end{array}
\right)
\left(
\begin{array}{c}
 W_{0}^{\nu}\\
 B^{\nu}
\end{array}
\right)
,
\end{align} 
where
\begin{align}
 \Pi_{W_3 W_3}
=&
 \Pi_{00} - \Pi_{01} (\Pi_{11})^{-1} \Pi_{10}
, \\ 
 \Pi_{W_3 B}
=&
 \Pi_{02} - \Pi_{01} (\Pi_{11})^{-1} \Pi_{12}
, \\ 
 \Pi_{B B}
=&
 \Pi_{22} - \Pi_{21} (\Pi_{11})^{-1} \Pi_{12}
.
\end{align}
Their explicit expressions are given in appendix~\ref{sec:STdetail}.
In a similar manner, we can calculate the charged sector, and we find
$\Pi_{W_1 W_1}(q^2)= \Pi_{W_3 W_3}(q^2)$. Therefore $\hat{T}=\hat{U}=0$
in this model. 
Using the definition given in Ref.~\cite{Barbieri:2004qk}, we find
\begin{align}
 \hat{S}
=&
 \frac{g_0^2 v_1^2 v_2^2}{g_1^2 ( v_1^2 + v_2^2)^2 + g_0^2 v_1^4}
, \\ 
 \hat{T}
=&
 0
, \\ 
 \hat{U}
=&
 0
, \\ 
W
=&
4 m_{W}^2 
\frac{g_0^2}{g_1^2}
\frac{1}{v_1^2 + v_2^2} 
\frac{v_1^4}{g_1^2 (v_1^2 + v_2^2)^2 + g_0^2 v_1^4}
, \\ 
Y
=&
4 m_{W}^2 
\frac{g_2^2}{g_1^2}
\frac{1}{v_1^2 + v_2^2} 
\frac{v_2^4}{g_1^2 (v_1^2 + v_2^2)^2 + g_2^2 v_2^4}
.
\end{align}
The central values, standard deviations, and correlations of these
parameters are given in table 4 in Ref.~\cite{Barbieri:2004qk}.
\begin{align}
 10^3 \hat{S}
=&
 0 \pm 1.3
\equiv 
 10^{3} \left( \hat{S}_0 \pm \sigma_{\hat{S}}\right)
, \\ 
 10^3 \hat{T}
=&
 0.1 \pm 0.9
\equiv 
 10^{3} \left( \hat{T}_0 \pm \sigma_{\hat{T}}\right)
, \\ 
 10^3 Y
=&
 0.1 \pm 1.2
\equiv 
 10^{3} \left( \hat{Y}_0 \pm \sigma_{\hat{Y}}\right)
, \\ 
 10^3 W
=&
 -0.4 \pm 0.8
\equiv 
 10^{3} \left( \hat{W}_0 \pm \sigma_{\hat{W}}\right)
, \\ 
\rho
=&
\left(
\begin{matrix}
 1    & 0.68 & 0.65 & -0.12 \\
 0.68 &    1 & 0.11 & 0.19 \\
 0.65 & 0.11 &   1  & -0.59 \\
-0.12 & 0.19 & -0.59& 1
\end{matrix}
\right)
.
\end{align}
The confidence ellipse is given as
\begin{align}
 \vec{v}^{T} V^{-1} \vec{v} = \sigma_{\text{CL}}^2
,
\end{align}
where
\begin{align}
 \vec{v}^{T}
=&
10^{3}
\left(
\begin{matrix}
\hat{S} - \hat{S}_0
&
\hat{T} - \hat{T}_0
&
Y- Y_0
&
W - W_0
\end{matrix}
\right)
, \\ 
V
=&
(10^{3})^2
\left(
\begin{matrix}
 \sigma_{\hat{S}} & 0 & 0 & 0 \\ 
 0 & \sigma_{\hat{T}} & 0 & 0 \\ 
 0 & 0 & \sigma_Y & 0 \\
 0 & 0 & 0 & \sigma_W \\
\end{matrix}
\right)
\rho
\left(
\begin{matrix}
 \sigma_{\hat{S}} & 0 & 0 & 0 \\ 
 0 & \sigma_{\hat{T}} & 0 & 0 \\ 
 0 & 0 & \sigma_Y & 0 \\
 0 & 0 & 0 & \sigma_W \\
\end{matrix}
\right)
,
\end{align}
and where
\begin{align}
 \sigma_{\text{CL}}^2 
=&
\begin{cases}
 & 4.71957 \quad (68.27 \text{\% CL})  \\
 & 7.77944 \quad (90 \text{\% CL})  \\
 & 9.48773 \quad (95 \text{\% CL})  \\
 & 13.2767 \quad (99 \text{\% CL})  \\
\end{cases}
.
\end{align}
The set of parameters should be in this ellipse.

There are three parameters relevant for the calculations, $g_1$, $r
\equiv v_2/ v_1$, and $v_3$. The rest of parameters such as $g_0$,
$g_2$, and $v_1$ (or $v_2$) are fixed so that $\alpha$, $M_Z$, and
$G_F$ are correctly reproduced.
Numerical results are shown in
Figs.~\ref{fig:constraint_(mwp,v3).eps}--\ref{fig:constraint_(mwp,r).eps},
where we take the mass of $W'$, $M_{W'}$, as the horizontal axis.
In Fig.~\ref{fig:constraint_(mwp,v3).eps}, we show excluded parameter
regions with fixed $r$, and those with fixed $v_3$ is shown in
Figs.~\ref{fig:constraint_(mwp,g1).eps}, and
\ref{fig:constraint_(mwp,r).eps}.
The regions to the left of the red lines are excluded from the
electroweak precision tests.
We see that the constraint from the electroweak precision measurements
is almost independent of the values of $g_1$ and $r$ but depend
on $v_3$. The lower bound on $m_{W'}$ is typically 1~TeV (2~TeV) for
$v_3 = 200$~GeV (100~GeV).

For $r=1$, corresponding to the parity conserving model for the
dynamical sector, one can see that the gauge coupling constant $g_1$
needs to be large such as $g_1 \sim 10$ in order to evade the
electroweak constraints (See
Fig.~\ref{fig:constraint_(mwp,v3).eps}). With such a large value, the
tree level analysis becomes not reliable. On the other hand, for $r \ll
1$ or $r \gg 1$, there can be consistent parameter regions with $g_1$
much smaller than $4 \pi$ (See Figs.~\ref{fig:constraint_(mwp,g1).eps}
and \ref{fig:constraint_(mwp,r).eps}).
This suggests that the dynamical sector is either parity violating, such
as chiral theories, or a theory which does not provide a particle
picture for the vector resonances, unlike the QCD.

\begin{figure}[tbp]
\begin{tabular}{cccc}
\begin{minipage}{0.24\hsize}
\begin{center}
 \includegraphics[angle=0, width=\hsize]{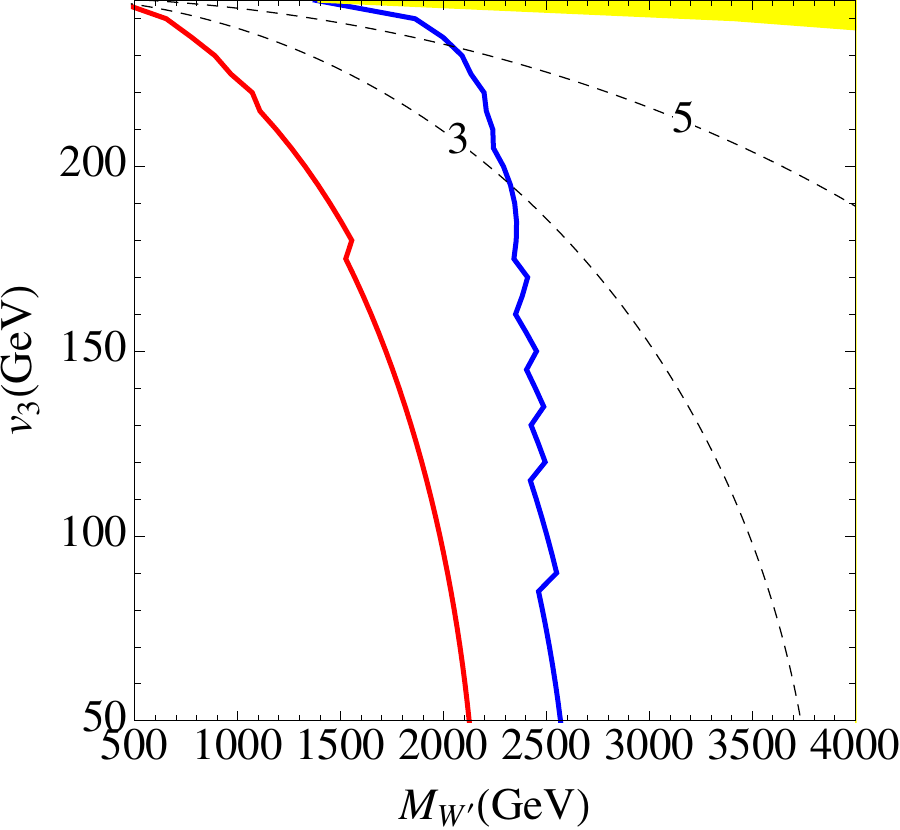}
\end{center}
\end{minipage}
\begin{minipage}{0.24\hsize}
\begin{center}
 \includegraphics[angle=0, width=\hsize]{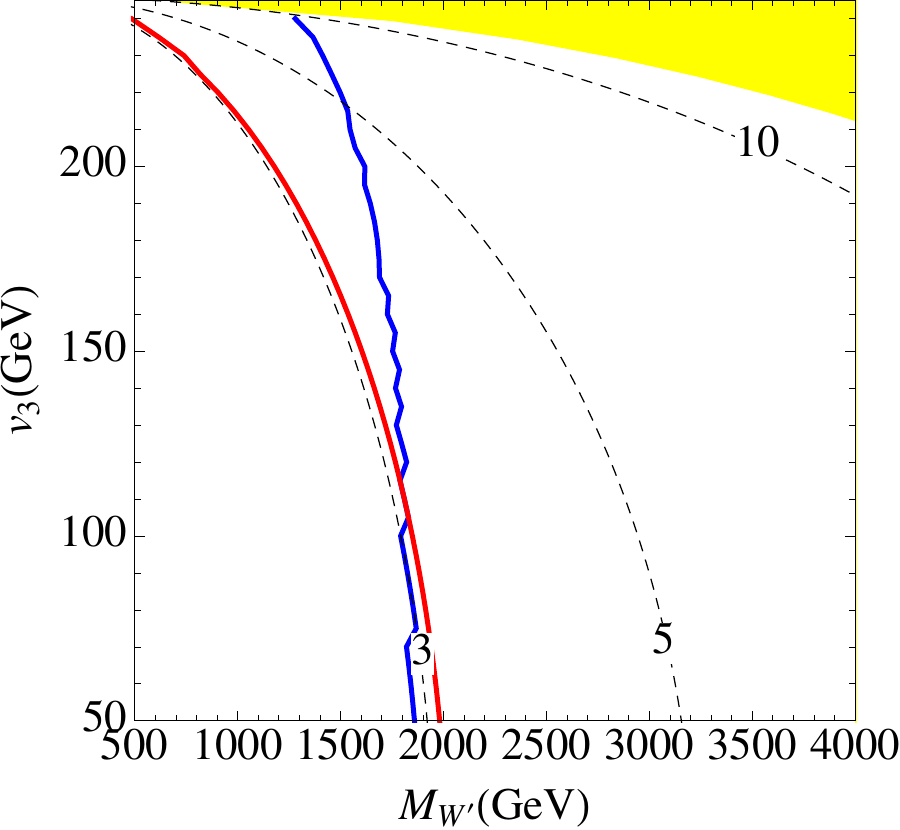}
\end{center}
\end{minipage}
\begin{minipage}{0.24\hsize}
\begin{center}
 \includegraphics[angle=0, width=\hsize]{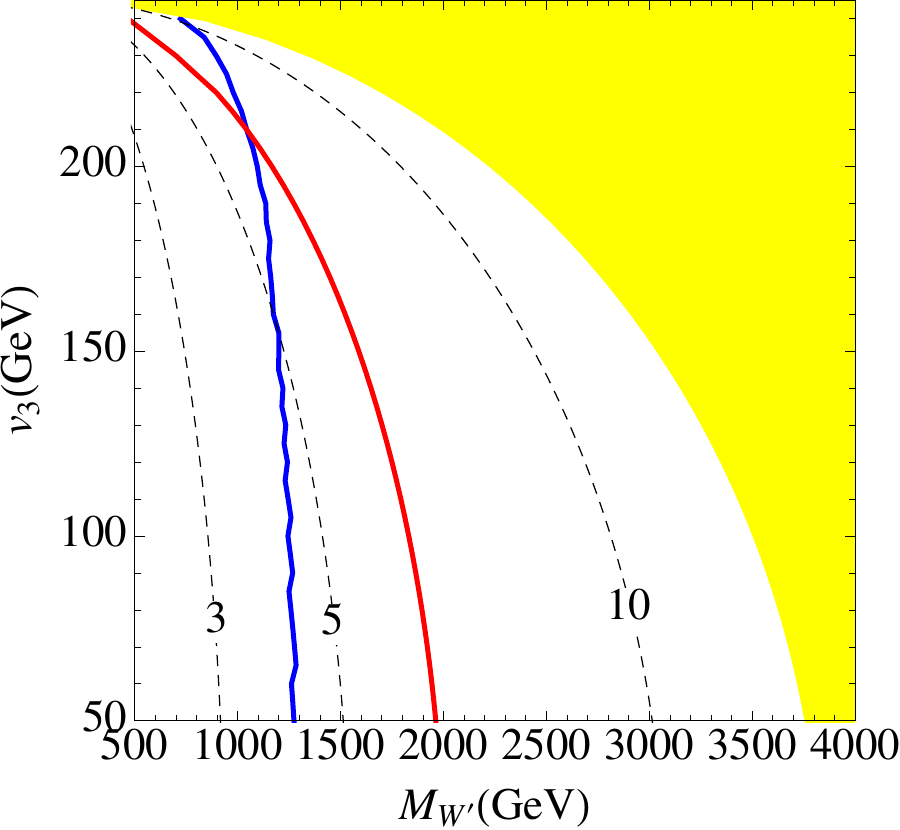}
\end{center}
\end{minipage}
\begin{minipage}{0.24\hsize}
\begin{center}
 \includegraphics[angle=0, width=\hsize]{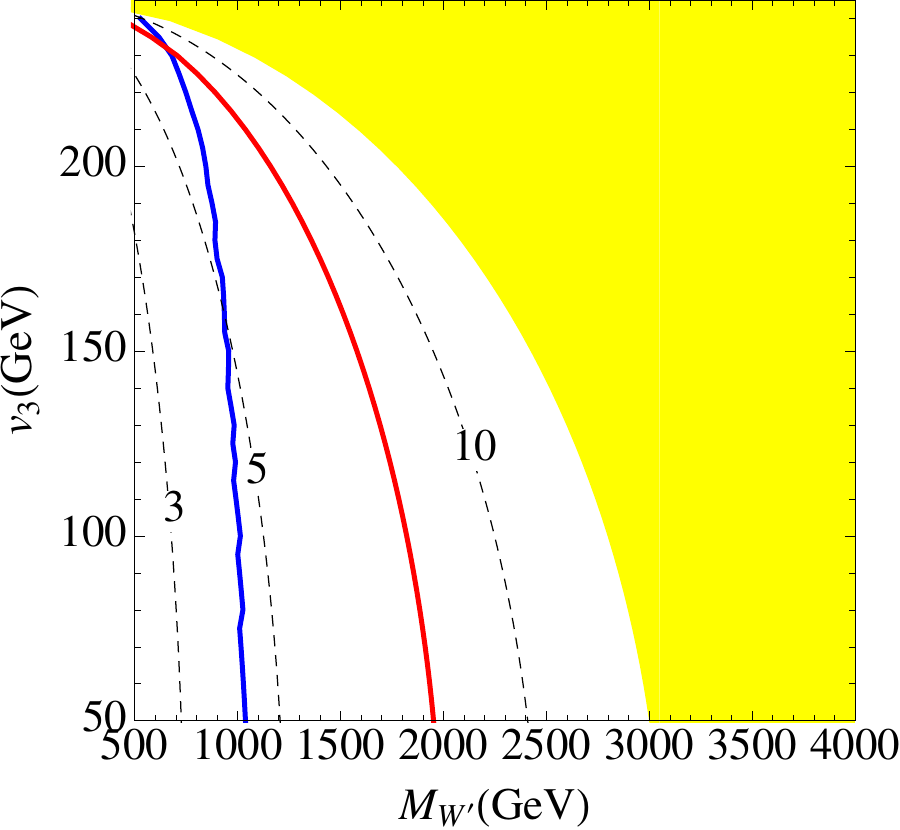}
\end{center}
\end{minipage}
\\ \\
\begin{minipage}{0.24\hsize}
\begin{center}
 \includegraphics[angle=0, width=\hsize]{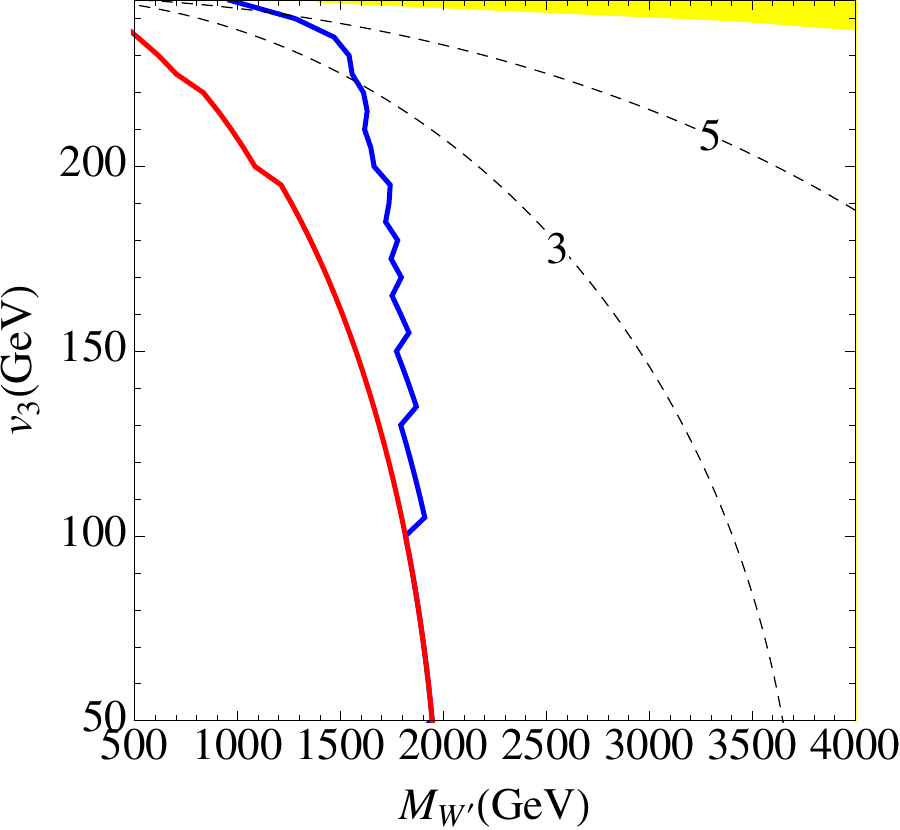}
\end{center}
\end{minipage}
\begin{minipage}{0.24\hsize}
\begin{center}
 \includegraphics[angle=0, width=\hsize]{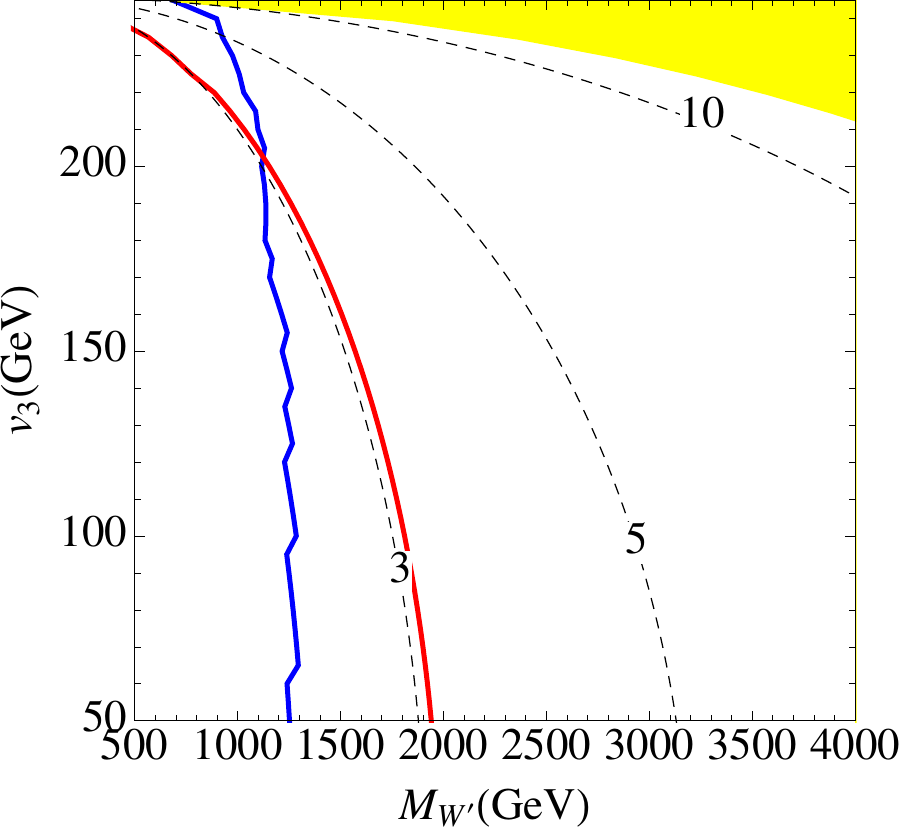}
\end{center}
\end{minipage}
\begin{minipage}{0.24\hsize}
\begin{center}
 \includegraphics[angle=0, width=\hsize]{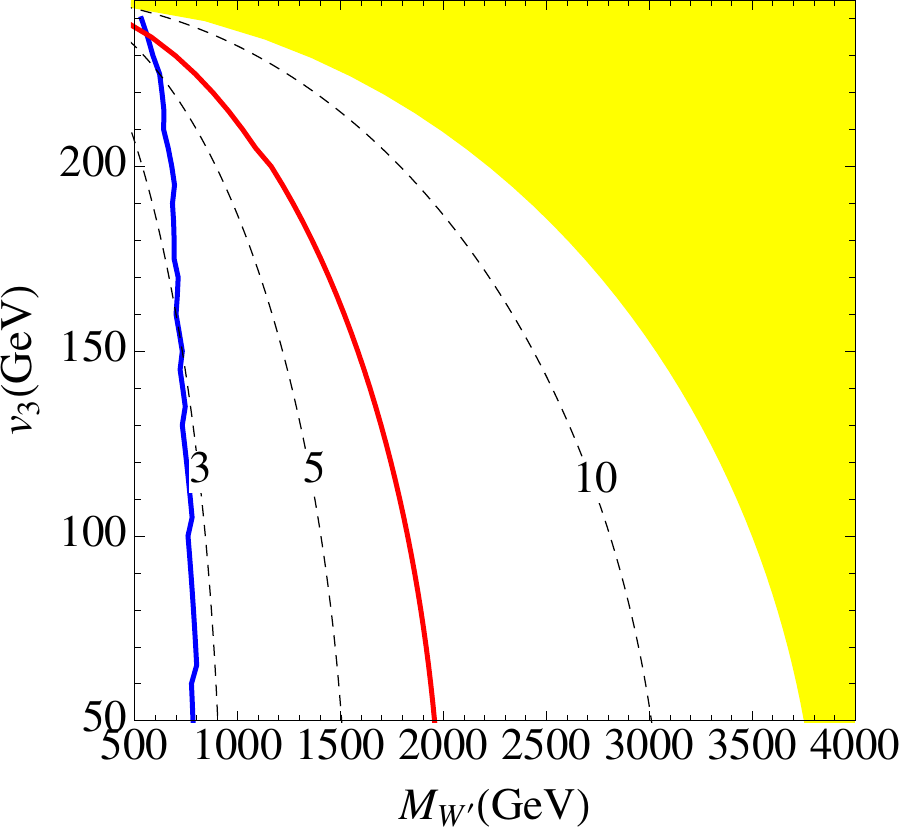}
\end{center}
\end{minipage}
\begin{minipage}{0.24\hsize}
\quad
\end{minipage}
\end{tabular}
 \caption{
Constraints in ($m_{W'}$, $v_3$)-plane.
 The left side of the red (blue) line is excluded by the
 electroweak precision measurements (the $W'/Z'$ search by the LHC). The
 numbers on the dashed lines are $g_1$ value. 
 The yellow region represents the region in which $g_1 \geq 4 \pi$. 
 In the first (second) row, we take $r$ = 0.1, 0.2, 0.5, 1 (10, 5, 2)
 from left to right column. 
}
\label{fig:constraint_(mwp,v3).eps}
 \end{figure}

\begin{figure}[tbp]
\begin{tabular}{cccc}
\begin{minipage}{0.24\hsize}
\begin{center}
 \includegraphics[angle=0, width=\hsize]{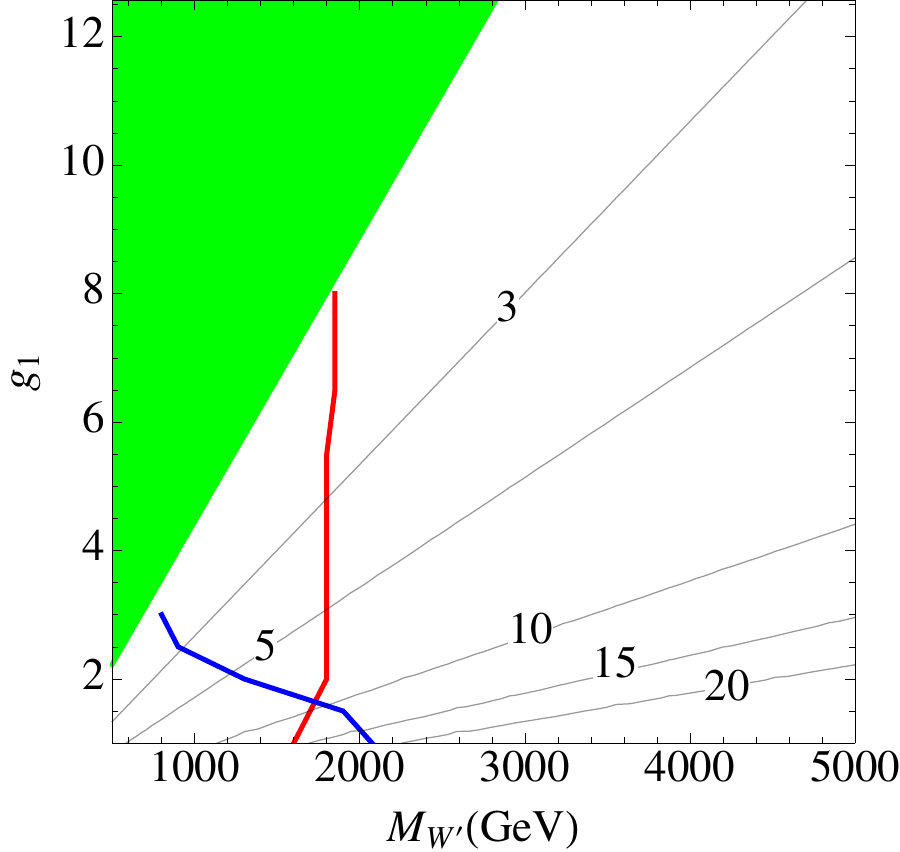}
\end{center}
\end{minipage}
\begin{minipage}{0.24\hsize}
\begin{center}
 \includegraphics[angle=0, width=\hsize]{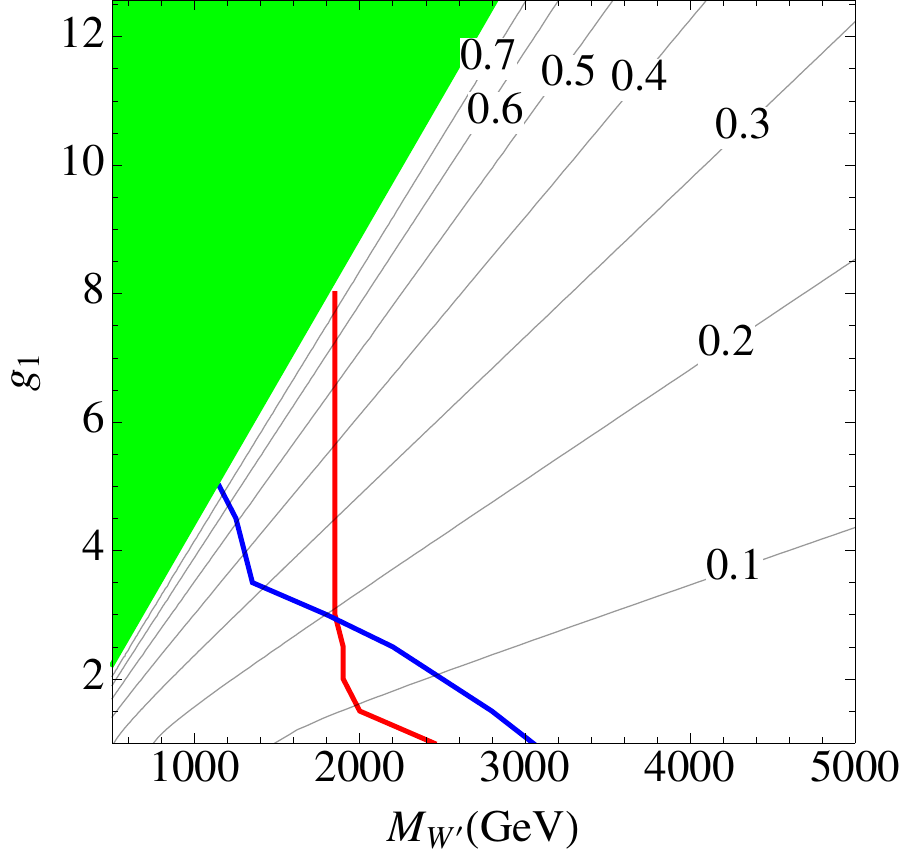}
\end{center}
\end{minipage}
\begin{minipage}{0.24\hsize}
\begin{center}
 \includegraphics[angle=0, width=\hsize]{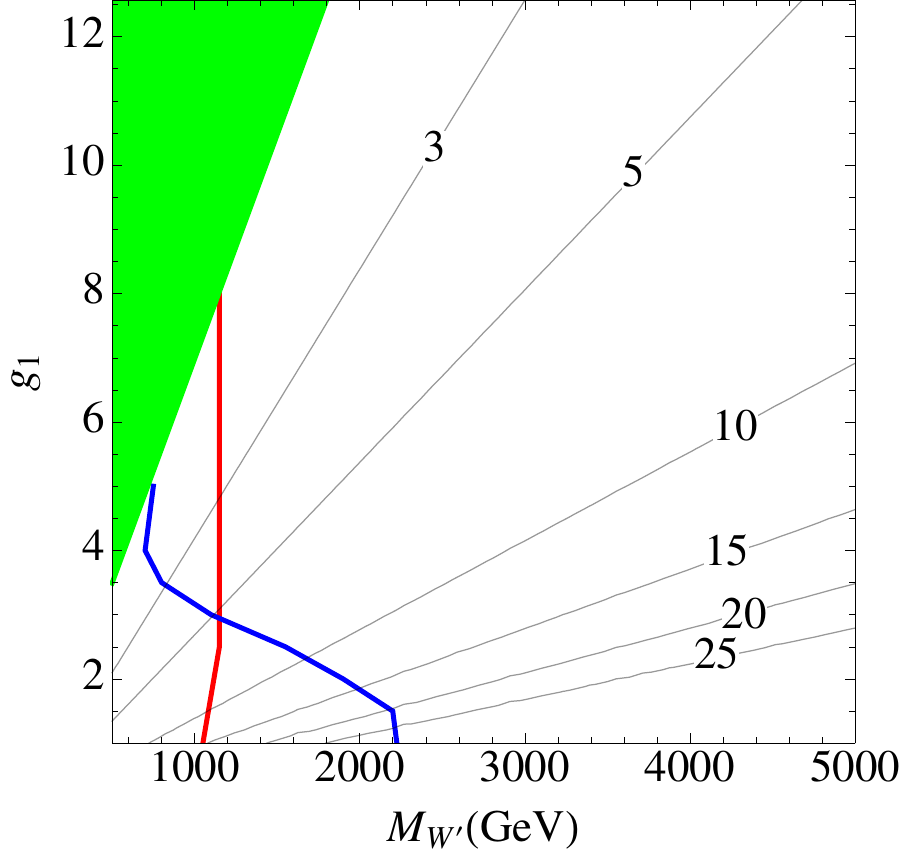}
\end{center}
\end{minipage}
\begin{minipage}{0.24\hsize}
\begin{center}
 \includegraphics[angle=0, width=\hsize]{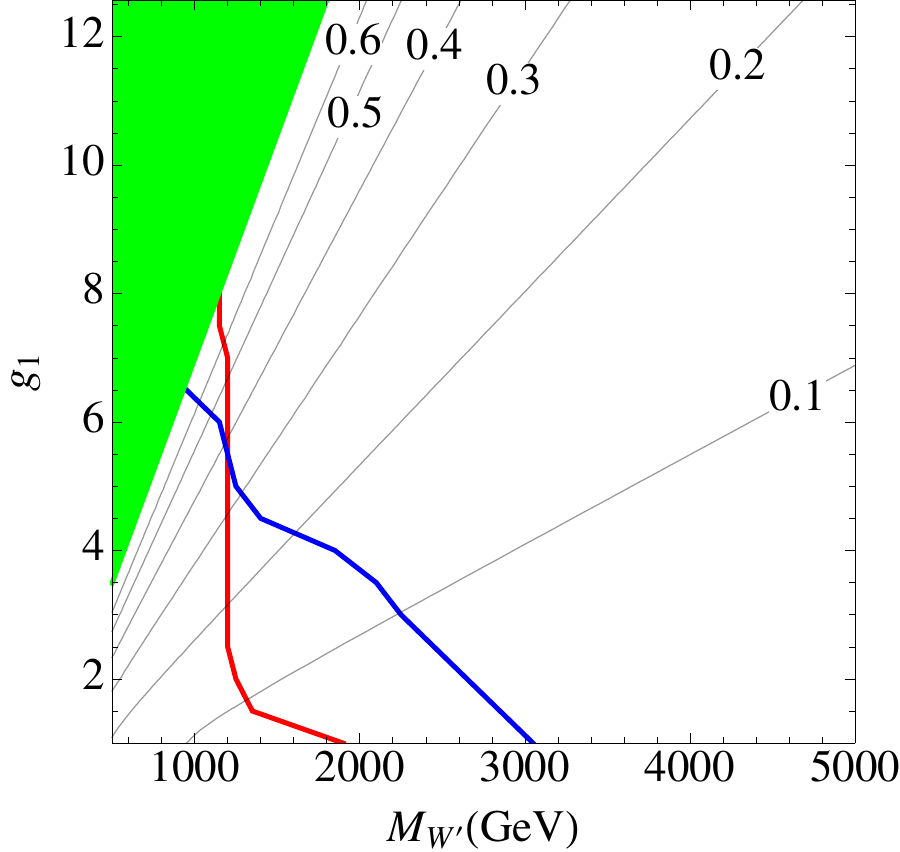}
\end{center}
\end{minipage}
\begin{minipage}{0.24\hsize}
\end{minipage}
\end{tabular}
 \caption{
Constraints in ($m_{W'}$, $g_1$)-plane. No physical solutions are in the
 left side of the green line, namely the gauge couplings and/or VEV's
 becomes complex numbers there.
 The left side of the red (blue) line is excluded by the
 electroweak precision measurements (the $W'/Z'$ search by the LHC). 
 From the left to right panels we take $v_3 = 100$~GeV ($r > 1$), $v_3 =
 100$~GeV ($r < 1$), $v_3 = 200$~GeV ($r > 1$), and $v_3 = 200$~GeV ($r < 1$)
}
\label{fig:constraint_(mwp,g1).eps}
 \end{figure}

\begin{figure}[tbp]
\begin{tabular}{cccc}
\begin{minipage}{0.24\hsize}
\begin{center}
 \includegraphics[angle=0, width=\hsize]{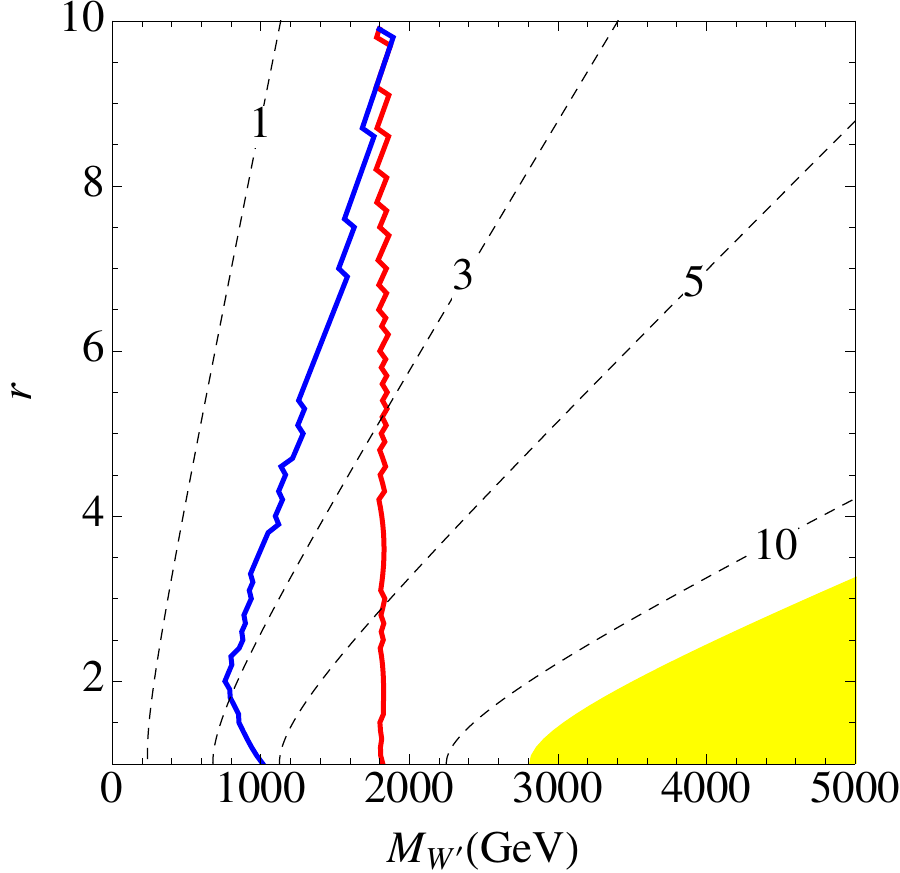}
\end{center}
\end{minipage}
\begin{minipage}{0.24\hsize}
\begin{center}
 \includegraphics[angle=0, width=\hsize]{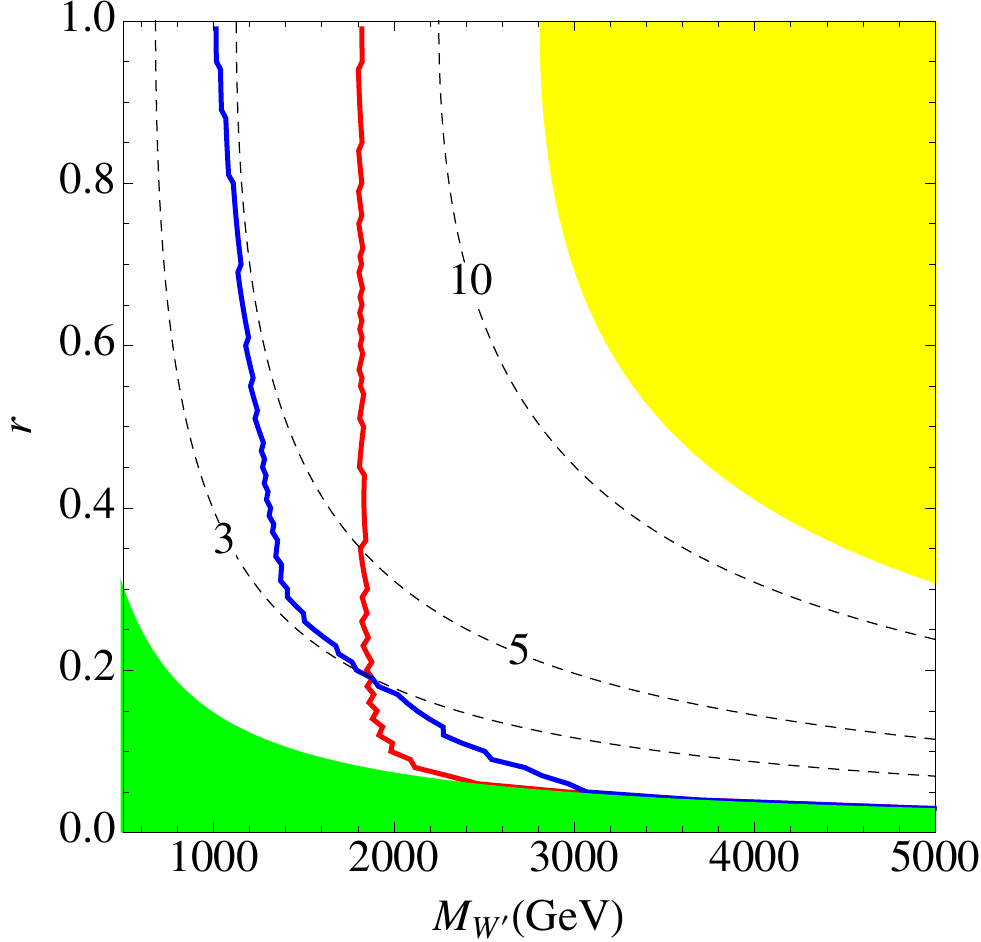}
\end{center}
\end{minipage}
\begin{minipage}{0.24\hsize}
\begin{center}
 \includegraphics[angle=0, width=\hsize]{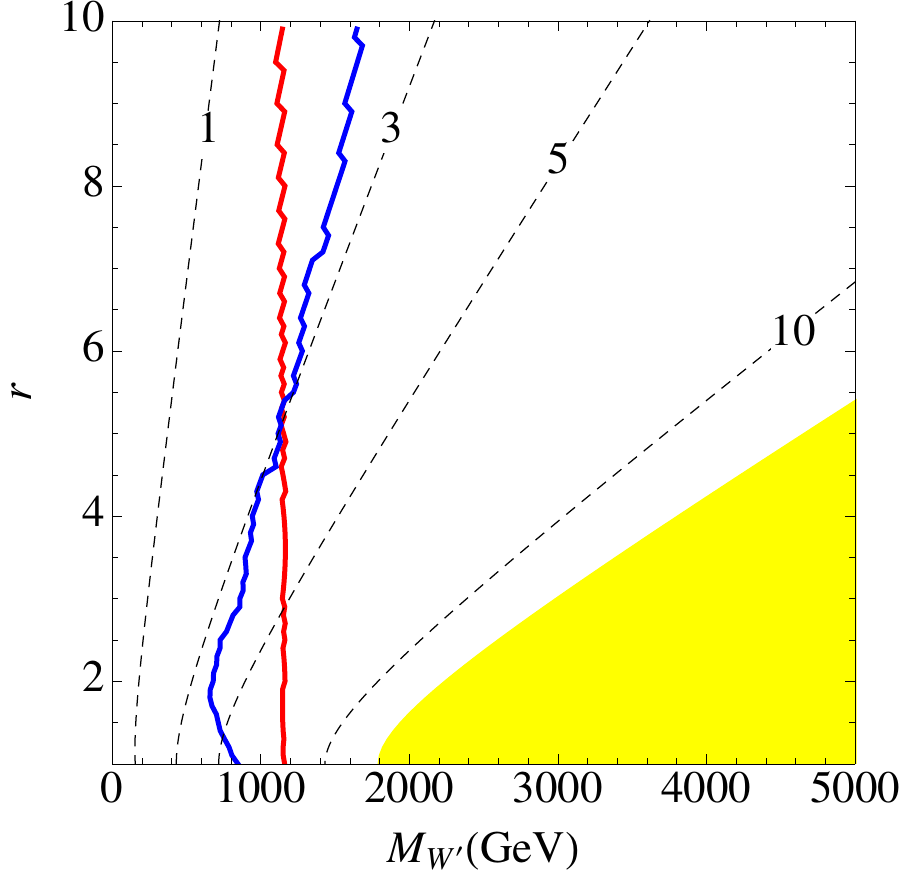}
\end{center}
\end{minipage}
\begin{minipage}{0.24\hsize}
\begin{center}
 \includegraphics[angle=0, width=\hsize]{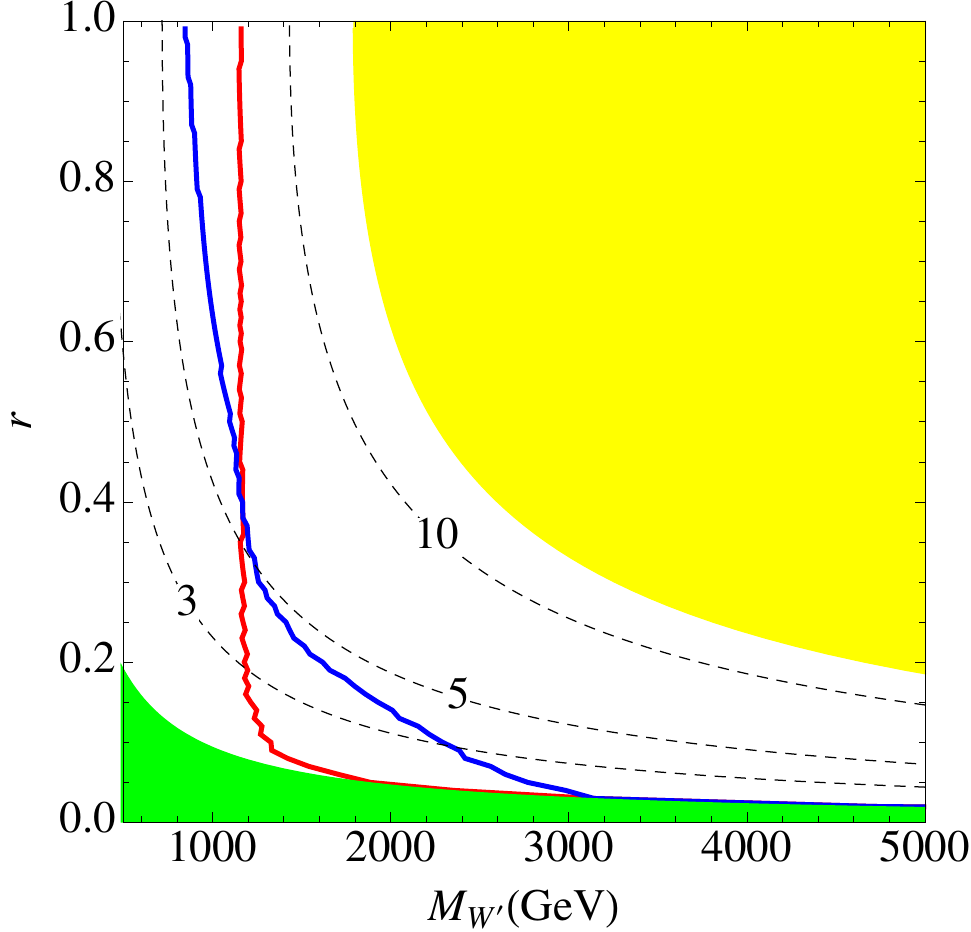}
\end{center}
\end{minipage}
\begin{minipage}{0.24\hsize}
\end{minipage}
\end{tabular}
 \caption{
Constraints in ($m_{W'}$, $r$)-plane. No physical solutions are in the
 the green region, namely the gauge couplings and/or VEV's
 becomes complex numbers there.
 The left side of the red (blue) line is excluded by the
 electroweak precision measurements (the $W'/Z'$ search by the LHC). 
 From the left to right panels we take $v_3 = 100$~GeV ($r > 1$), $v_3 =
 100$~GeV ($r < 1$), $v_3 = 200$~GeV ($r > 1$), and $v_3 = 200$~GeV ($r < 1$)
}
\label{fig:constraint_(mwp,r).eps}
 \end{figure}

\subsection{Constraints from direct searches for $W'$ and $Z'$ at the
  LHC} \label{sec:directDetection} 

In this section, we discuss the
bounds from the direct searches for $W'$ and $Z'$ bosons at the LHC
experiments.
Both ATLAS and CMS groups provides bounds on the combinations $\sigma
\cdot $Br for each decay modes as a function of the mass of the $W'$ and
$Z'$.  Since there are couplings to fermions through the mixing with
Standard Model gauge bosons, $W'$ and $Z'$ can be produced via Drell-Yan
processes. 
If the leptonic decay modes, namely $W' \to \ell \nu$ and $Z' \to \ell
\ell$, have sizable branching fraction, there are quite strong
bounds. Whereas when $W'$ is almost fermiophobic, its main decay mode is
$W' \to WZ$. We, therefore, consider constraints from both processes:
$pp \to W' \to WZ$, $pp \to W' \to \ell \nu$, and $pp \to Z' \to \ell
\ell$.

\begin{figure}[tbp]
\begin{tabular}{cc}
\begin{minipage}{0.5\hsize}
\begin{center}
 \includegraphics[angle=0, width=\hsize]{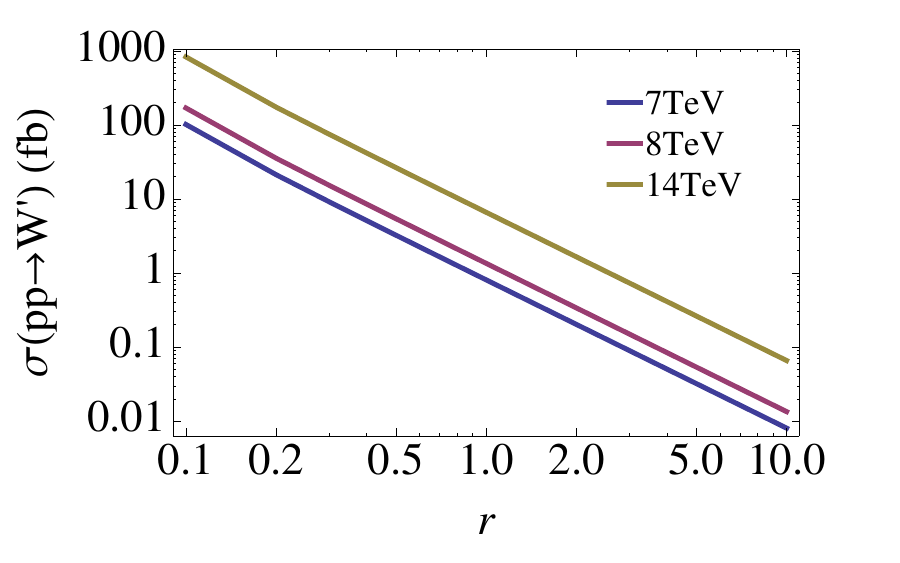}
\end{center}
\end{minipage}
\begin{minipage}{0.5\hsize}
\begin{center}
 \includegraphics[angle=0, width=\hsize]{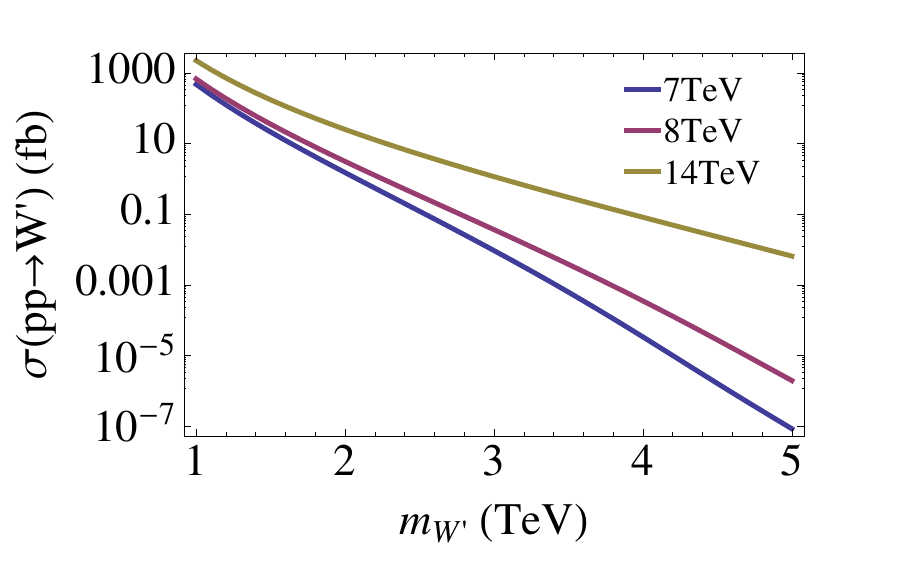}
\end{center}
\end{minipage}
\\ \\
\begin{minipage}{0.5\hsize}
\begin{center}
  \includegraphics[angle=0, width=\hsize]{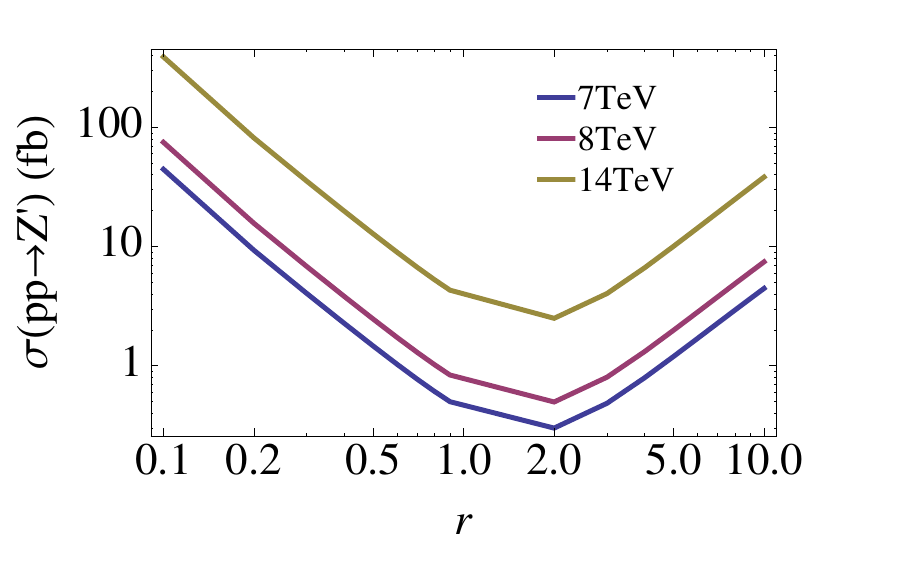}
\end{center}
\end{minipage}
\begin{minipage}{0.5\hsize}
\begin{center}
  \includegraphics[angle=0, width=\hsize]{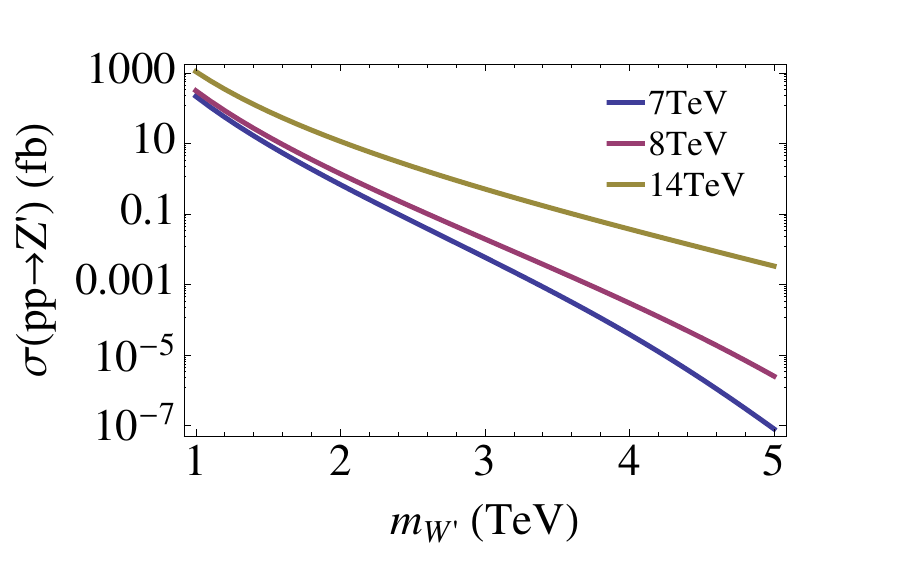}
\end{center}
\end{minipage}
\end{tabular}
 \caption{
The production cross section of $W'$ and $Z'$ via Drell-Yan process as
 functions of $r$, and $m_{W'}$. We take $m_W' =1500$~GeV and
 $v_3=200$~GeV in the left column, $v_3=200$~GeV and $r=0.2$ in the
 right column. Here $\sigma(pp \to W') = \sigma(pp \to W'^{-}) +
 \sigma(pp \to W'^{+})$. Here we use CTEQ6 for PDFs~\cite{Pumplin:2002vw}.
}
\label{fig:xsec.eps}
 \end{figure}

We here give some qualitative discussion.
The production cross sections are proportional to couplings squared,
\begin{align}
 \sigma(q \bar{q} \to Z')
\propto&
( g_{Z'ffL}^2 + g_{Z'ffR}^2) 
\\ 
\simeq &
\frac{e^2}{s_Z^2}
\frac{m_W^2}{m_{W'}^2}
\frac{1}{r^2}
\left(
1 - \frac{v_3^2}{v^2}
\right)
\left[
\left(
\left(
1
-
r^2 \frac{s_Z^2}{c_Z^2}
\right)
T^3
+
r^2
\frac{s_Z^2}{c_Z^2}
Q
\right)^2
+
\left(
r^2
\frac{s_Z^2}{c_Z^2}
Q
\right)^2
\right]
, \\ 
 \sigma(q \bar{q'} \to W')
\propto&
g_{W'ff}^2
\\ 
\simeq &
\frac{e^2}{s_Z^2}
\frac{m_W^2}{m_{W'}^2}
\frac{1}{r^2}
\left(
1 - \frac{v_3^2}{v^2}
\right)
.
\end{align}
In the large $r$ region, productions of $W'$ are suppressed but $Z'$ is
enhanced. Hence $Z'$, rather than $W'$, is expected to give stronger
bound on parameter space in the large $r$ region. On the other hand,
both give similar bounds in small $r$ region.  Notice that this
difference between $W'$ and $Z'$ is due to large breaking of the
custodial symmetry as we discussed in Sec.~\ref{sec:model}.
We show the cross sections of $W'$ and $Z'$ via Drell-Yan production at
LHC in Fig.~\ref{fig:xsec.eps}. The $r$ dependence discussed here is now
apparent in the left column in this figure.

\begin{figure}[tbp]
\begin{tabular}{cccc}
\begin{minipage}{0.24\hsize}
\begin{center}
 \includegraphics[angle=0, width=\hsize]{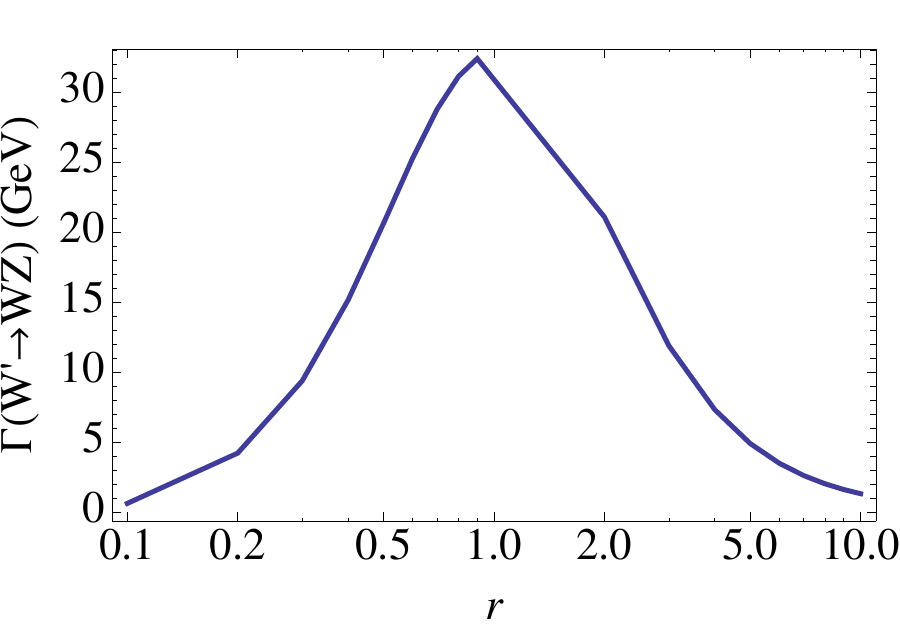}
\end{center}
\end{minipage}
\begin{minipage}{0.24\hsize}
\begin{center}
 \includegraphics[angle=0, width=\hsize]{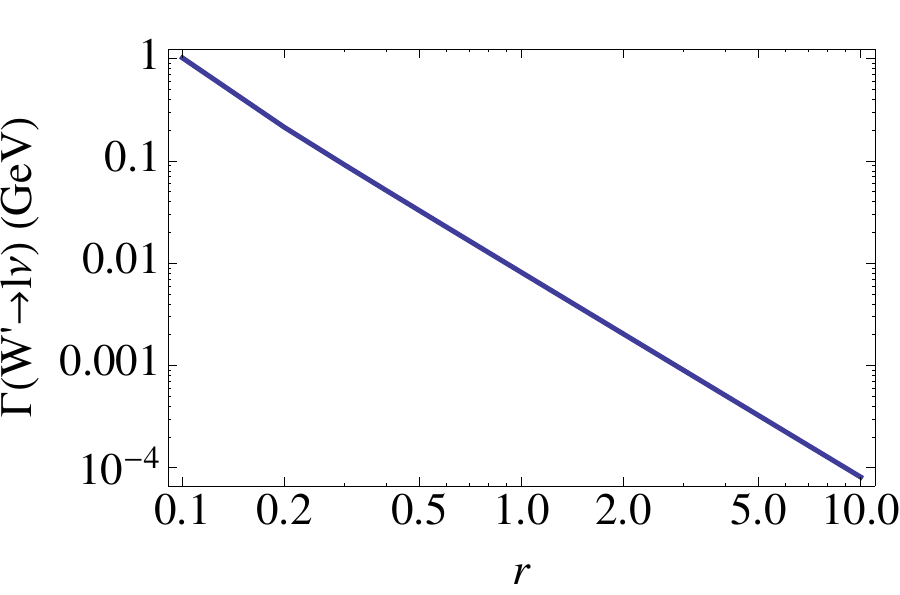}
\end{center}
\end{minipage}
\begin{minipage}{0.24\hsize}
\begin{center}
 \includegraphics[angle=0, width=\hsize]{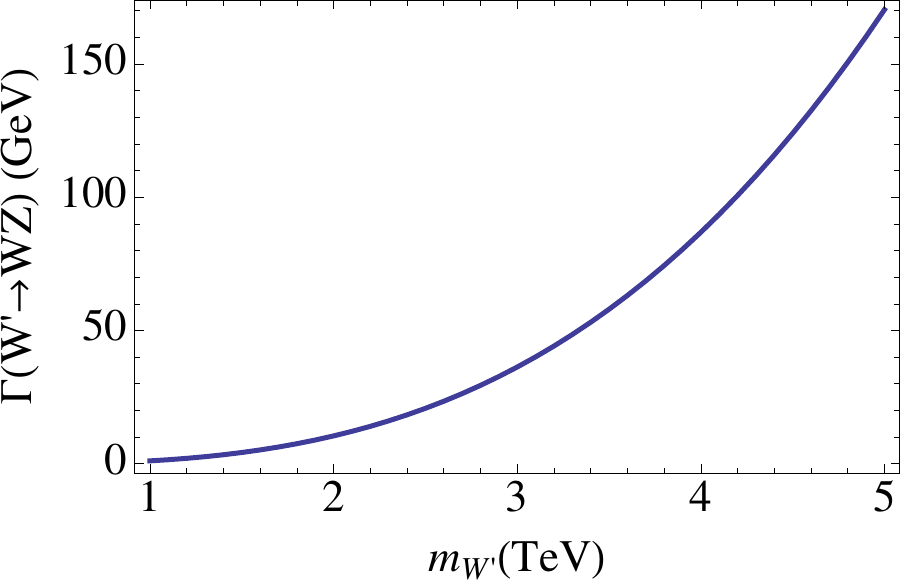}
\end{center}
\end{minipage}
\begin{minipage}{0.24\hsize}
\begin{center}
 \includegraphics[angle=0, width=\hsize]{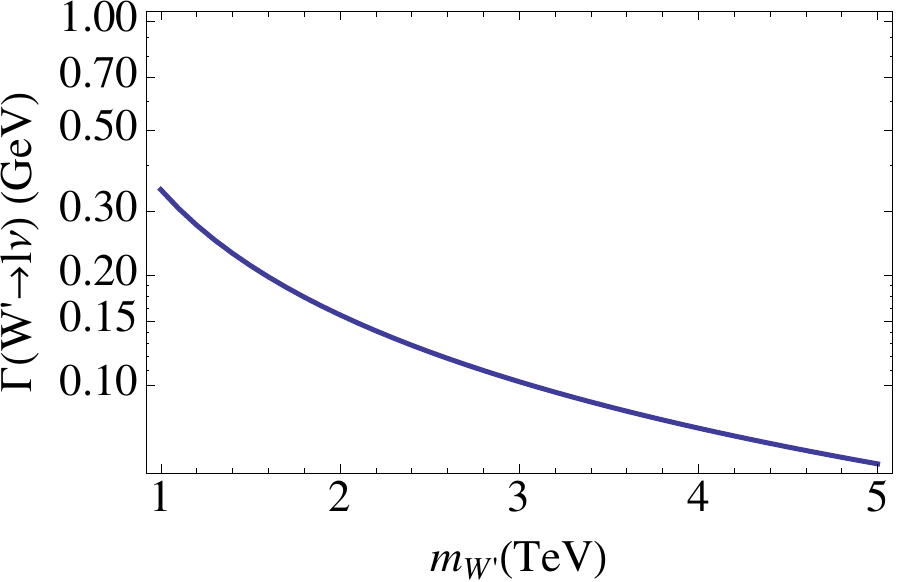}
\end{center}
\end{minipage}
\\ \\
\begin{minipage}{0.24\hsize}
\begin{center}
 \includegraphics[angle=0, width=\hsize]{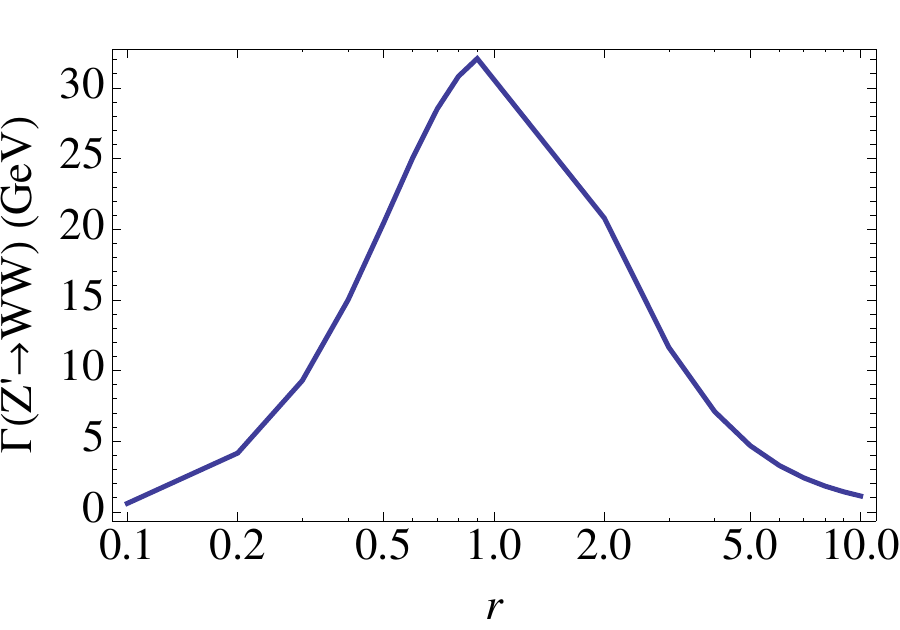}
\end{center}
\end{minipage}
\begin{minipage}{0.24\hsize}
\begin{center}
 \includegraphics[angle=0, width=\hsize]{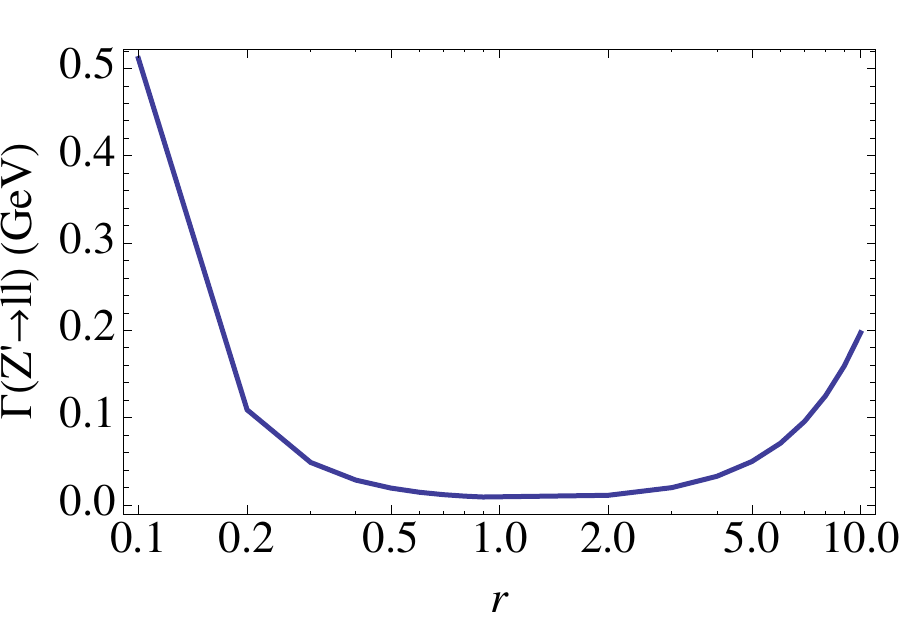}
\end{center}
\end{minipage}
\begin{minipage}{0.24\hsize}
\begin{center}
 \includegraphics[angle=0, width=\hsize]{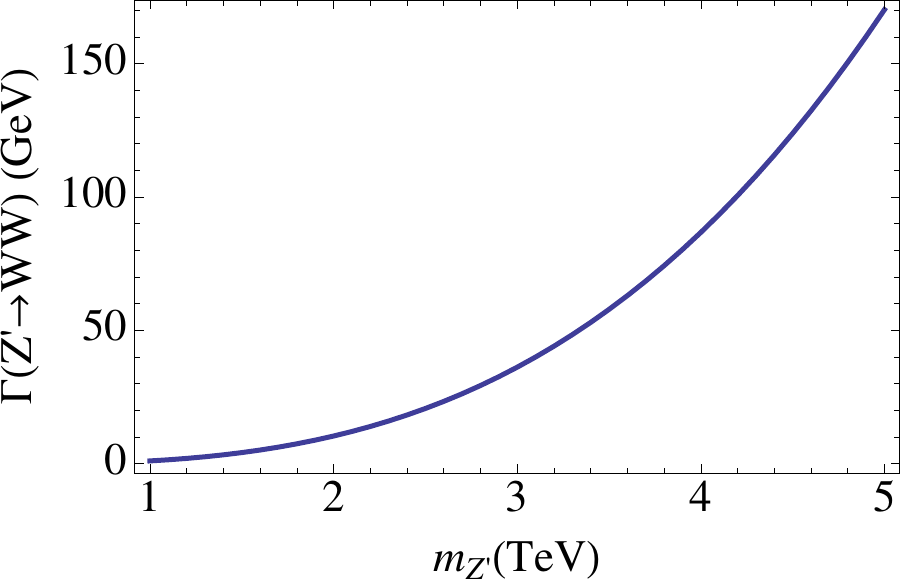}
\end{center}
\end{minipage}
\begin{minipage}{0.24\hsize}
\begin{center}
 \includegraphics[angle=0, width=\hsize]{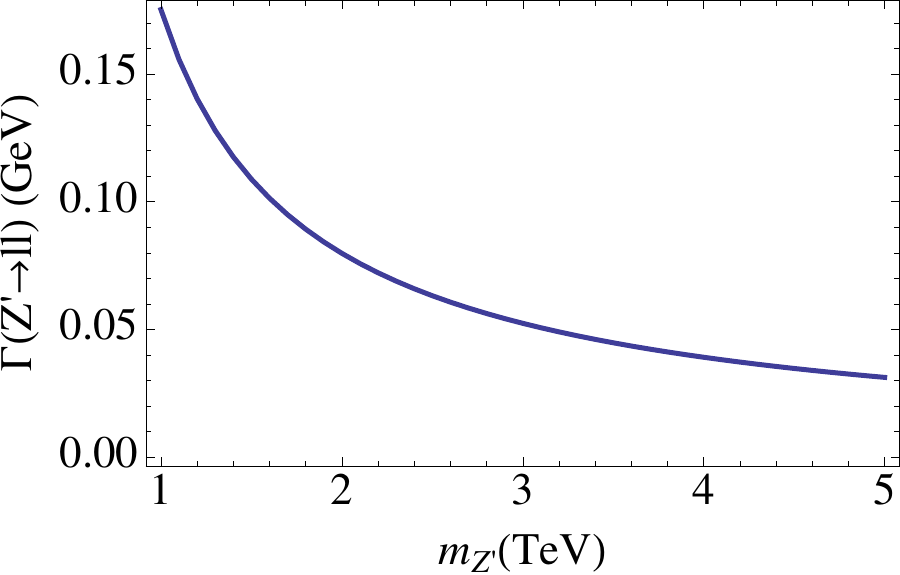}
\end{center}
\end{minipage}
\end{tabular}
 \caption{
 The partial decay widths of $W'$ and $Z'$ as functions of $r$ and $m_{W'}$.
 We take $m_{W'} =1500$~GeV and $v_3=200$~GeV in the left four panels,
 $v_3=200$~GeV and $r=0.2$ in the right four panels. Here $l\nu = e\nu +
 \mu \nu$, and $ll = ee + \mu \mu$.
}
\label{fig:W'Z'widths}
 \end{figure}

The partial decay widths of $W'$ and $Z'$ are
\begin{align}
\Gamma(Z' \to WW)
\simeq&
\frac{1}{48 \pi}
\frac{m_{W'}^3}{v^2}
\frac{r^2}{(1+r^2)^2}
\left(
1 - \frac{v_3^2}{v^2}
\right)
, \\ 
\Gamma(Z' \to Zh)
\simeq&
\frac{1}{48\pi}
\frac{m_{W'}^3}{v^2}
\frac{r^2}{(1+r^2)^{3}}
\left(
-
r
w_h^{1}
+
w_h^{2}
\right)^2
, \\ 
\Gamma(Z' \to f \bar{f'})
\simeq&
\frac{1}{24 \pi}
N_c
\frac{m_{W}^2}{m_{W'}}
\frac{e^2}{s_Z^2}
\frac{1}{r^2}
\left(
1 - \frac{v_3^2}{v^2}
\right)
\left(
\left(
\left(
1
-
r^2 \frac{s_Z^2}{c_Z^2}
\right)
T^3
+
r^2
\frac{s_Z^2}{c_Z^2}
Q
\right)^2
+
\left(
r^2
\frac{s_Z^2}{c_Z^2}
Q
\right)^2
\right)
, \\ 
\Gamma(W' \to WZ)
\simeq&
\frac{1}{48 \pi}
\frac{m_{W'}^3}{v^2}
\frac{r^2}{(1+r^2)^2}
\left(
1 - \frac{v_3^2}{v^2}
\right)
, \\ 
\Gamma(W' \to Wh)
\simeq&
\frac{1}{48\pi}
\frac{m_{W'}^3}{v^2}
\frac{r^2}{(1+r^2)^{3}}
\left(
-
r
w_h^{1}
+
w_h^{2}
\right)^2
, \\ 
\Gamma(W' \to f \bar{f'})
\simeq&
\frac{1}{48 \pi}
N_c
\frac{m_{W}^2}{m_{W'}}
\frac{e^2}{s_Z^2}
\frac{1}{r^2}
\left(
1 - \frac{v_3^2}{v^2}
\right)
.
\end{align}
Here we keep leading terms in the $(m_{W}/m_{W'})$ expansion. We find that
$\Gamma(Z' \to WW) \simeq \Gamma(W' \to WZ)$ and
$\Gamma(Z' \to Zh) \simeq \Gamma(W' \to Wh)$ in this approximation. 
To see which of bosonic and fermionic decay modes is more important, we take
their ratio:
\begin{align}
 \frac{\Gamma(W' \to WZ)}{\Gamma(W' \to f \bar{f}')}
\simeq&
 \frac{1}{4 N_c}
 \frac{m_{W'}^4}{m_W^4}
 \frac{r^4}{(1+r^2)^2}
.
\label{eq:W'2WZ_vs_W'2ff}
\end{align}
We see that bosonic decay mode is dominant except the small $r$ region.
The $Z'$ case is similar to Eq.~(\ref{eq:W'2WZ_vs_W'2ff}) in the small
$r$ region, but it has extra $r^{-2}$ in the large $r$ region. Then $Z'
\to f\bar{f}'$ as well as $Z' \to WW$ is important in the large $r$ region.
We plot partial decay widths in Fig.~\ref{fig:W'Z'widths}. The
qualitative features discussed here are explicit as one can see in this
figure.

Now we calculate $\sigma \cdot$Br for $W'$ and $Z'$, and compare the
results from the searches at the LHC.  We use the bounds on $pp \to W'
\to WZ$~\cite{ATLAS8_WZ, CMS7_WZ_3lv}, $pp \to W' \to \ell
\nu$~\cite{ATLAS7_lv, CMS8_lv}, and $pp \to Z' \to \ell
\ell$~\cite{ATLAS8_ll, CMS8_ll}.
In this section, we restrict ourselves to consider the parameter space
in which Eq.~(\ref{eq:partially_compositeness}) is satisfied.
Then we can omit $V' \to Vh$
process. We also omit some other channels including heavier Higgs bosons
and/or charged scalars, which highly depend on parameters in the Higgs
potential. After taking into account these processes, the
constraints might be weaker because they change the total decay
width. 

The numerical results are shown in
Figs.~\ref{fig:constraint_(mwp,v3).eps}--\ref{fig:constraint_(mwp,r).eps}
as blue lines.
In a large parameter space, the electroweak precision test gives
stronger bound. The exceptions are regions with large and small $r$,
that is where $g_1$ can be small.
Since the $W'/Z'$ to fermion couplings are induced by the gauge boson
mixings of order $g_0/g_1$, the production and decay rates are enhanced
for a small $g_1$.
In these regions, the LHC experiments is starting to give stronger
bounds than the electroweak precision tests (see
Fig.~\ref{fig:constraint_(mwp,g1).eps}).

\section{Signal strength of 125~GeV Higgs}
The lightest Higgs boson $h$ is a mixture of $H_1$, $H_2$, and $H_3$,
and thus the properties are modified from the Standard Model
predictions. We here discuss the production/decay properties of $h$.
We start off by calculating ratios of partial decay widths. In the
processes which exist at tree level, they are given as the ratio of the
corresponding couplings given in Sec.~\ref{sec:couplings};
\begin{align}
 \frac{\Gamma(h \to ff)}{\Gamma(h \to ff)_{\text{SM}}}
=&
 \kappa_f^2
, \quad
 \frac{\Gamma(h \to WW)}{\Gamma(h \to WW)_{\text{SM}}}
=
 \kappa_W^2
, \quad
 \frac{\Gamma(h \to ZZ)}{\Gamma(h \to ZZ)_{\text{SM}}}
=
 \kappa_Z^2
.
\end{align}
We also define $\kappa$'s for the loop induced processes:
\begin{align}
 \kappa_g^2
\equiv &
 \frac{\Gamma(h \to gg)}{\Gamma(h \to gg)_{\text{SM}}} 
=
 \frac{\sigma(gg \to h)}{\sigma(gg \to h)_{\text{SM}}} 
, \\ 
 \kappa_\gamma^2
\equiv &
 \frac{\Gamma(h \to \gamma \gamma)}{\Gamma(h \to \gamma \gamma)_{\text{SM}}} 
.
\end{align}
The diagrams which contribute to $\kappa_g^2$ are the same as those in
the Standard Model. The only difference from the Standard Model is the
$h$ couplings to the SM fermions which is given as $\kappa_f$. Hence
\begin{align}
 \kappa_g^2
=&
 \kappa_f^2
.
\end{align} 
$\kappa_\gamma$ is more complicated because $W'$ and $H^{\pm}$
contribute to the process as well. The partial decay width for $h \to
\gamma \gamma$ is given as
\begin{align}
\Gamma(h \to \gamma \gamma)
\simeq&
m_h^3
\frac{\alpha_{em}^2}{16\pi^3}
\sqrt{2} G_F
\left| 
\frac{1}{3}
Q_t^2
N_c
\kappa_f
-
2.1
\kappa_{W}
-
\frac{7}{4}
\kappa_{W'}
+
\frac{1}{12}
\kappa_{H^{\pm}}
\right|^2
.
\end{align}
Here we take $m_h \simeq 125$~GeV.
Then we have
\begin{align}
\kappa_\gamma^2
\simeq&
\left| 
\left(
0.27
\frac{v}{v_3} 
-
1.3
\frac{v_3}{v}
\right)
w_h^{3}
+
0.005
\kappa_{H^{\pm}}
-
1.3
\frac{1}{(1+r^2)^{3/2}}
\sqrt{1 - \frac{v_3^2}{v^2}}
\left(
r^3
w_h^{1}
+
w_h^{2}
\right)
\right. \nonumber \\ & 
\quad \quad
\left.
-
1.1
\frac{r}{(1+r^2)^{3/2}}
\frac{1}{\sqrt{1-\frac{v_3^2}{v^2}}}
\left(
w_h^{1}
+
r
w_h^{2}
\right)
\right|^2
.
\label{eq:kappa_gamma}
\end{align}
There are four terms in Eq.~(\ref{eq:kappa_gamma}): The first term
consists of top quark contribution and a part of $W$ 
contributions. The second terms is the charged scalar contribution.
The third term is a part of $W$ contributions. The fourth term is
the $W'$ contribution.
We take $v_3$ as same order as $v$ to keep the perturbativity of the
Yukawa coupling. Then we find that the charged scalar contribution is
negligible. 
As long as we take partially compositeness condition in
Eq.~(\ref{eq:partially_compositeness}), 
the third term is negligible. On the other hand, the fourth
term can be visible because its denominator becomes small for
$v_3 \sim v$. Since the fourth term highly depends on $r$,
$r$ dependence of $\kappa_\gamma$ is large.
We can calculate signal strengths by using $\kappa$'s.
\begin{align}
 \mu(gg \to h \to X)
=&
\frac{ \kappa_g^2 \kappa_X^2}
{ \kappa_f^2 \text{Br}_{h\to ff}^{\text{SM}}  
+ \kappa_W^2  \text{Br}_{h\to WW}^{\text{SM}}  
+ \kappa_Z^2  \text{Br}_{h\to ZZ}^{\text{SM}}  
+ \kappa_g^2  \text{Br}_{h\to gg}^{\text{SM}}  
+ \kappa_\gamma^2  \text{Br}_{h\to \gamma \gamma}^{\text{SM}}  
+ \kappa_{Z \gamma}^2 \text{Br}_{h\to Z \gamma}^{\text{SM}}  
}
\\ 
\simeq&
\frac{ \frac{v^2}{v_3^2}(w_h^3)^2 }
{ \frac{3}{4} \frac{v^2}{v_3^2}(w_h^3)^2
  +
  \frac{1}{4} 
\left(
\frac{r^3}{(1+r^2)^{3/2}}
\sqrt{1 - \frac{v_3^2}{v^2}}
w_h^{1}
+
\frac{1}{(1+r^2)^{3/2}}
\sqrt{1 - \frac{v_3^2}{v^2}}
w_h^{2}
+
\frac{v_3}{v}
w_h^{3}
\right)^2
}
\kappa_X^2
.
\label{eq:mu_gg2h2X}
\end{align}
Here we calculate Br$_{h \to X}^{\text{SM}}$ with $m_h = 125$~GeV.
Since we take $v_3 \sim v$ and $|w_h^3|^2 \gg |w_h^1|^2, |w_h^2|^2$, $r$
dependence is only in $\kappa_X$. Therefore, the $r$ dependences of
$\kappa_\gamma$, $\kappa_f$, and $\kappa_{W/Z}$ are large, absent, and
weak, respectively. Another important feature is the $m_{W'}$
dependence. We find that $m_{W'}$ dependence is absent at the leading
order. These features are shown in Fig.~\ref{fig:mu(mwp,r).eps}.  In
this figure, we plot the signal strengths in ($m_{W'}, r$)-plane with
$w_h^1 = 0.1, w_h^2 = 0.5$. In this $w_h^1$ and $w_h^2$ choice, $h$ is
$\sim 30\%$ composite because $|w_h^1|^2 + |w_h^2|^2 \simeq
0.3$. These calculations are performed numerically 
and we do not use the approximated formulae given in this section.

We also show the signal strengths on ($w_h^1$, $w_h^2$)-plane.  We take
$v_3 = 200$~GeV, $m_{W'} = 2500$~GeV, and $r = 0.1$ in
Fig.~\ref{fig:mu(vh1,vh2)_200_2500_0.1.eps}.\footnote{In this parameter
point, $g_1$ is larger than $1$ but still smaller than $\sqrt{4 \pi}$.}
We find that $w_h^1 \sim w_h^2 \sim 0$ region is disfavored. One of
them, $w_h^2$ in this example, should take sizable value. This means
that the lightest Higgs boson has to have a component of not only the
elementary sector ($H_3$) but also the composite sector ($H_1$ and
$H_2$), namely Higgs boson needs to be partially composite.
\begin{figure}[tbp]
\begin{tabular}{cccc}
\begin{minipage}{0.24\hsize}
\begin{center}
 \includegraphics[width=\hsize, angle=0]{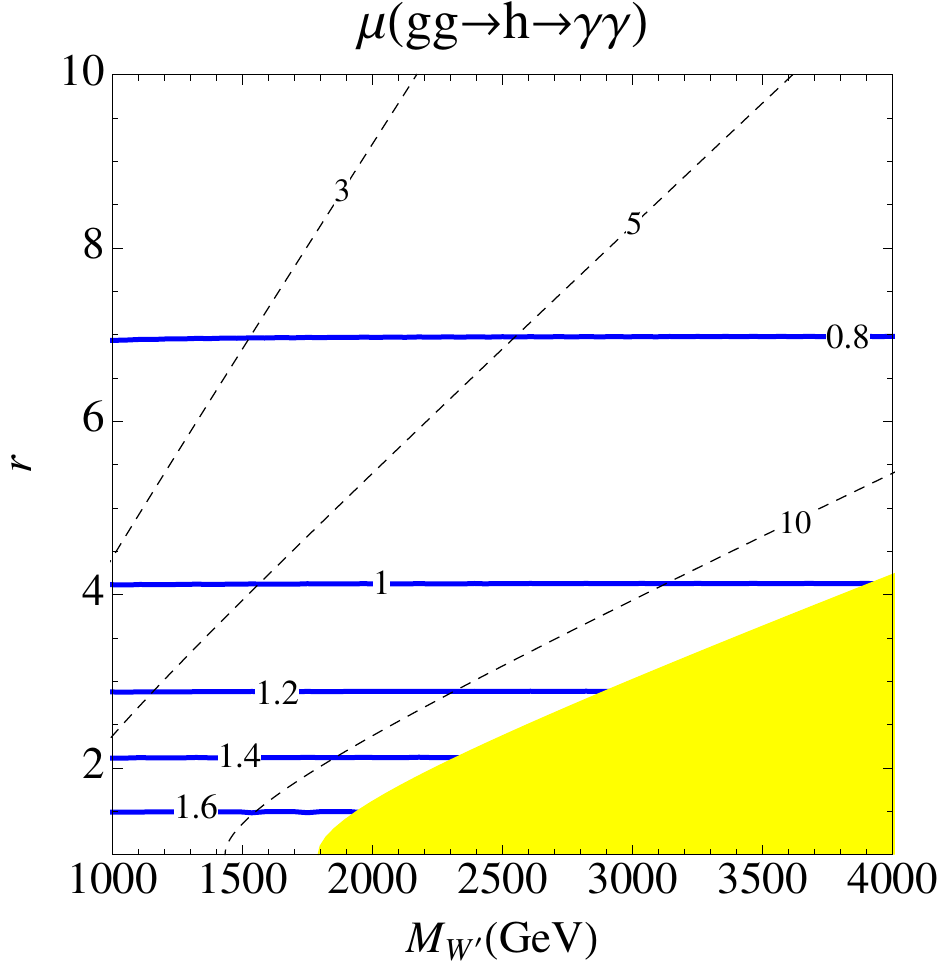}
\end{center}
\end{minipage}
\begin{minipage}{0.24\hsize}
\begin{center}
 \includegraphics[width=\hsize, angle=0]{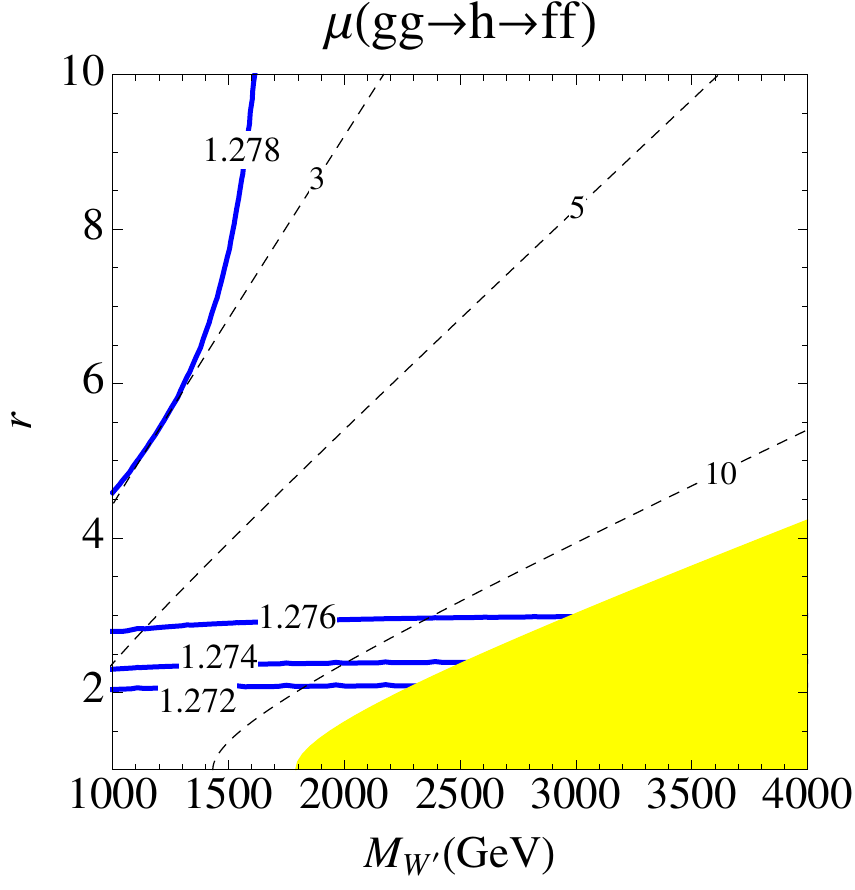}
\end{center}
\end{minipage}
\begin{minipage}{0.24\hsize}
\begin{center}
 \includegraphics[width=\hsize, angle=0]{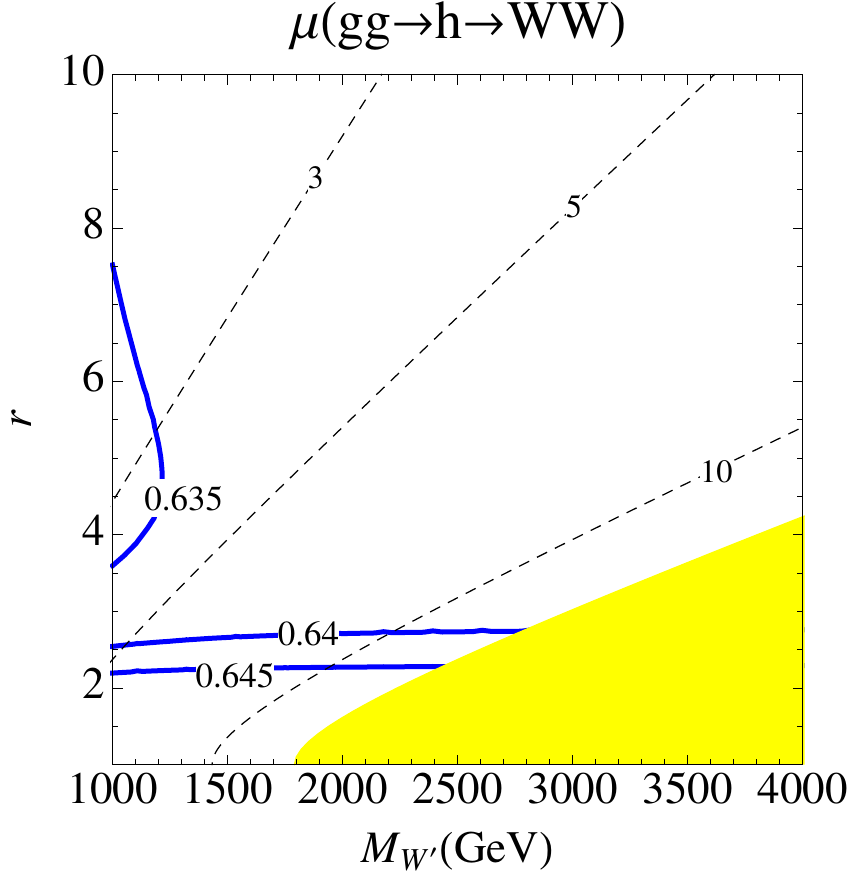}
\end{center}
\end{minipage}
\begin{minipage}{0.24\hsize}
\begin{center}
 \includegraphics[width=\hsize, angle=0]{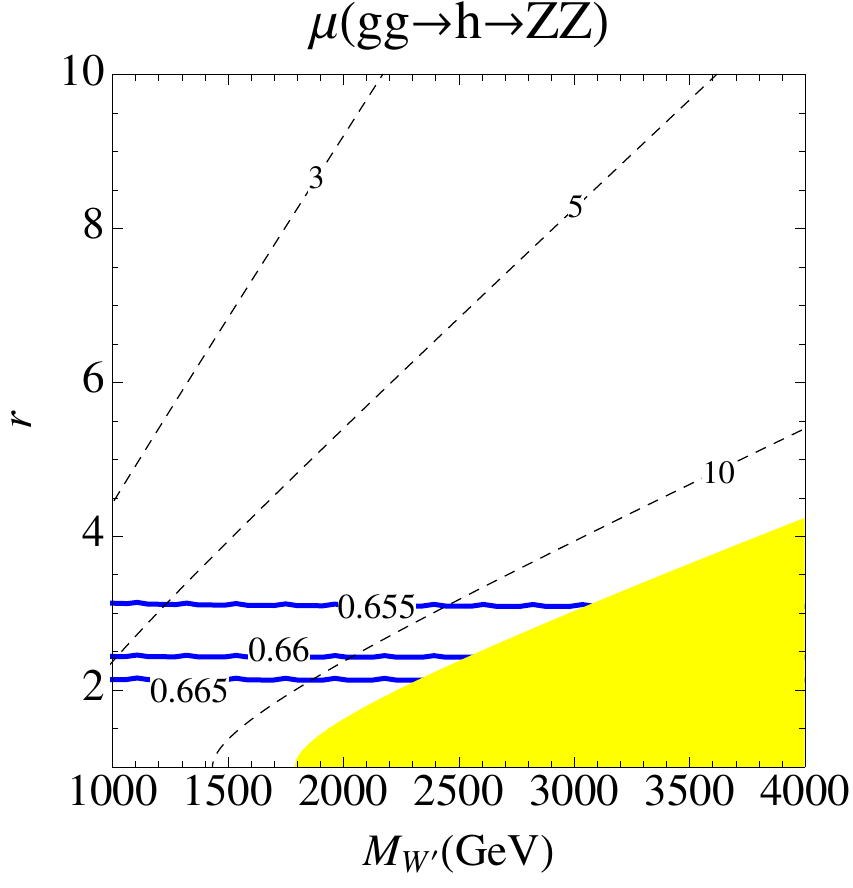}
\end{center}
\end{minipage}
\\
\\
\begin{minipage}{0.24\hsize}
\begin{center}
 \includegraphics[width=\hsize, angle=0]{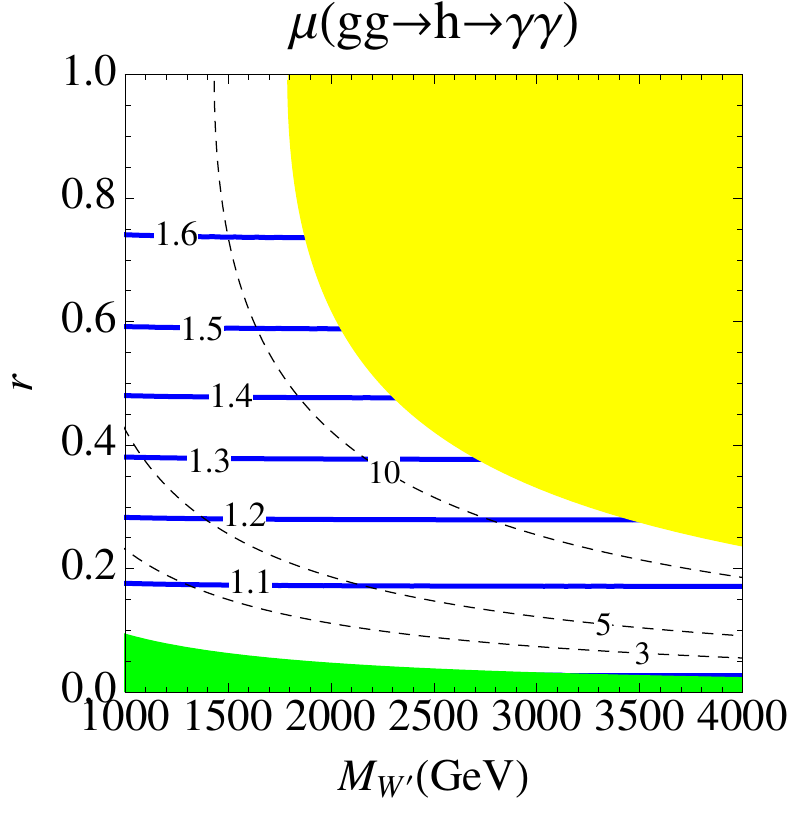}
\end{center}
\end{minipage}
\begin{minipage}{0.24\hsize}
\begin{center}
 \includegraphics[width=\hsize, angle=0]{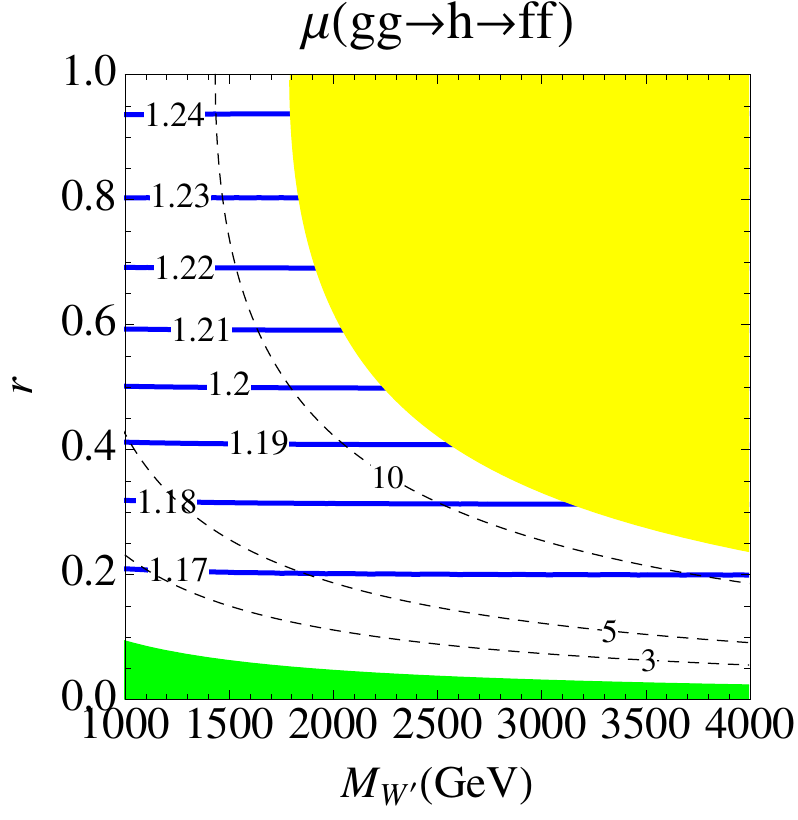}
\end{center}
\end{minipage}
\begin{minipage}{0.24\hsize}
\begin{center}
 \includegraphics[width=\hsize, angle=0]{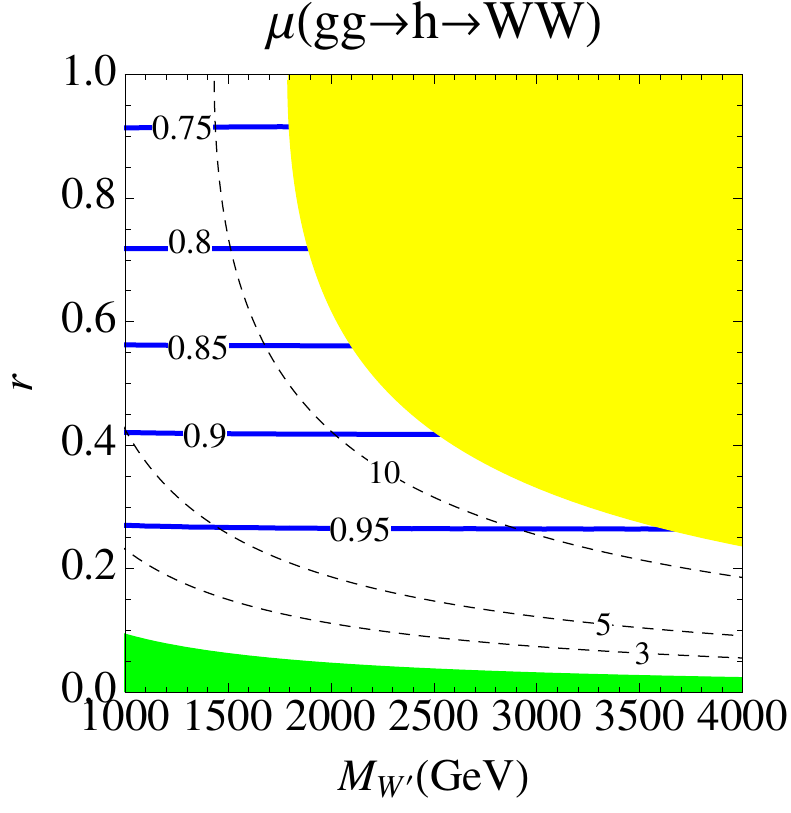}
\end{center}
\end{minipage}
\begin{minipage}{0.24\hsize}
\begin{center}
 \includegraphics[width=\hsize, angle=0]{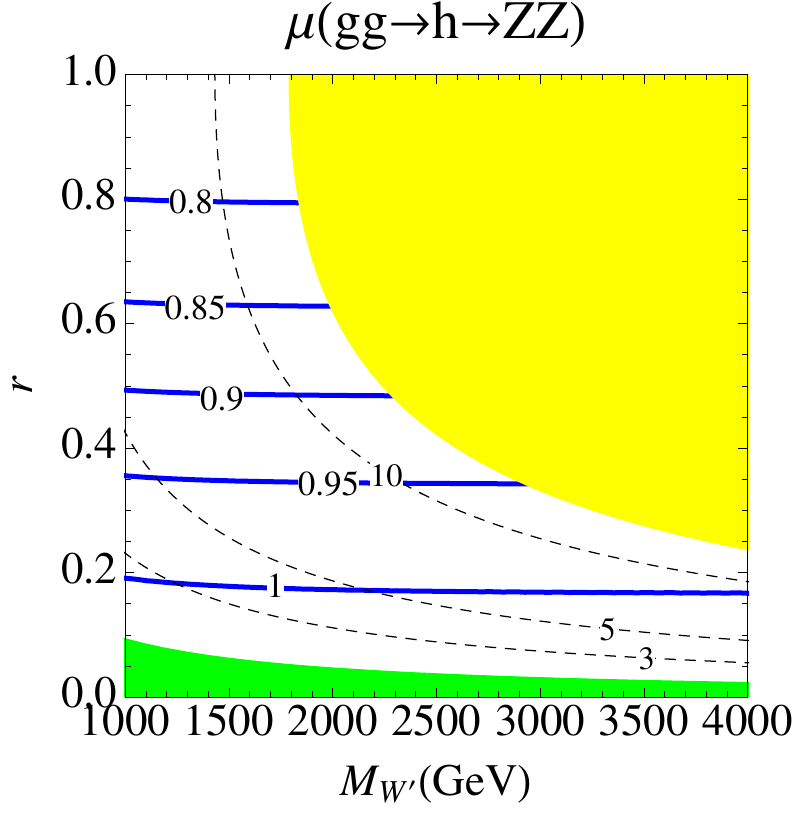}
\end{center}
\end{minipage}
\end{tabular}
 \caption{Signal strengths on the ($M_{W'}$, $r$)-plane. 
 The blue lines and dashed lines represent the signal strengths and
 $g_1$ values respectively.
 The vertical axis is the ratio of the two VEVs, $r = v_2/v_1$. 
 The yellow region stands for the region in which $g_1 > 4 \pi$.
 In the the green region, the gauge couplings and/or VEV's
 becomes complex numbers.
 The upper (lower) column show the $r>1 \ (r<1)$ region.
 The parameter choices here are $w_h^1 = 0.1$, $w_h^2 = 0.5$, and $v_3 =
 200$~GeV. 
}
\label{fig:mu(mwp,r).eps}
 \end{figure} 
\begin{figure}[tbp]
\begin{tabular}{cccc}
\begin{minipage}{0.24\hsize}
\begin{center}
 \includegraphics[width=\hsize, angle=0]{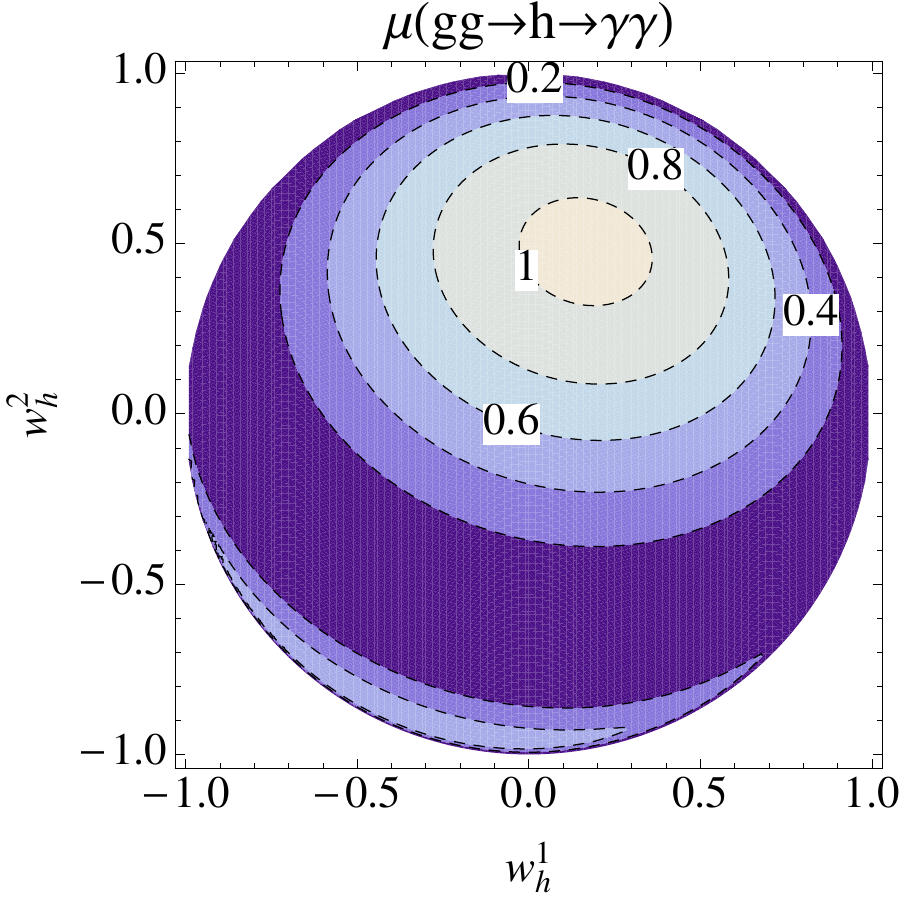}
\end{center}
\end{minipage}
\begin{minipage}{0.24\hsize}
\begin{center}
 \includegraphics[width=\hsize, angle=0]{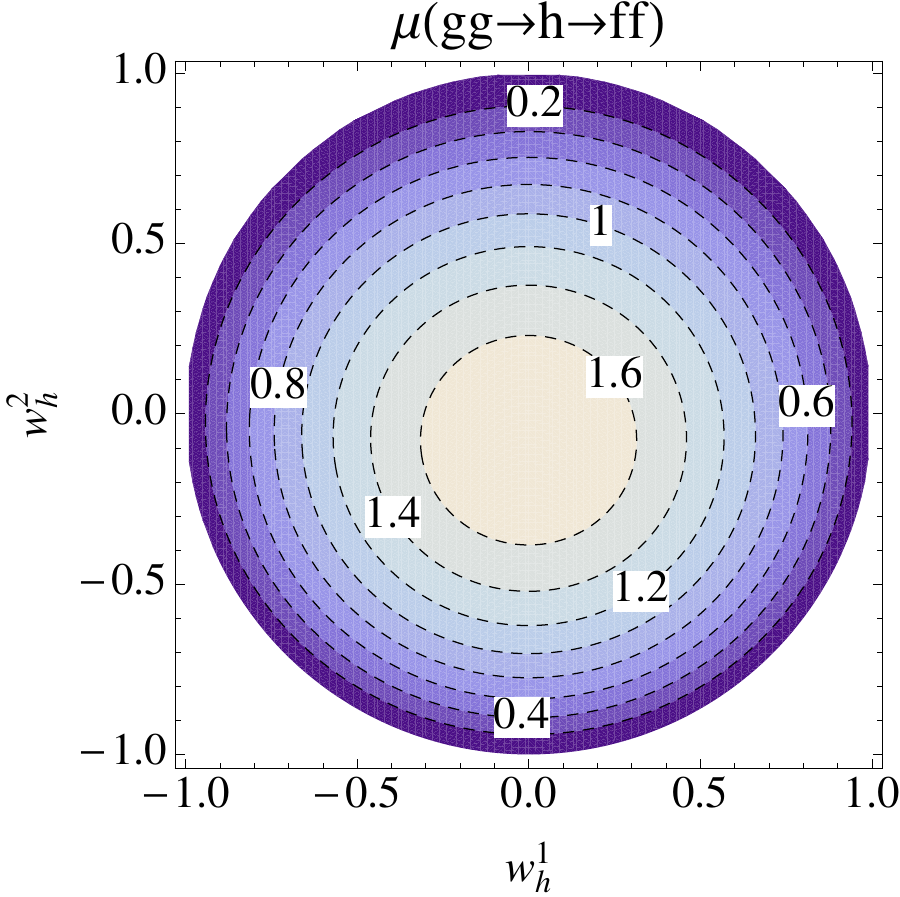}
\end{center}
\end{minipage}
\begin{minipage}{0.24\hsize}
\begin{center}
 \includegraphics[width=\hsize, angle=0]{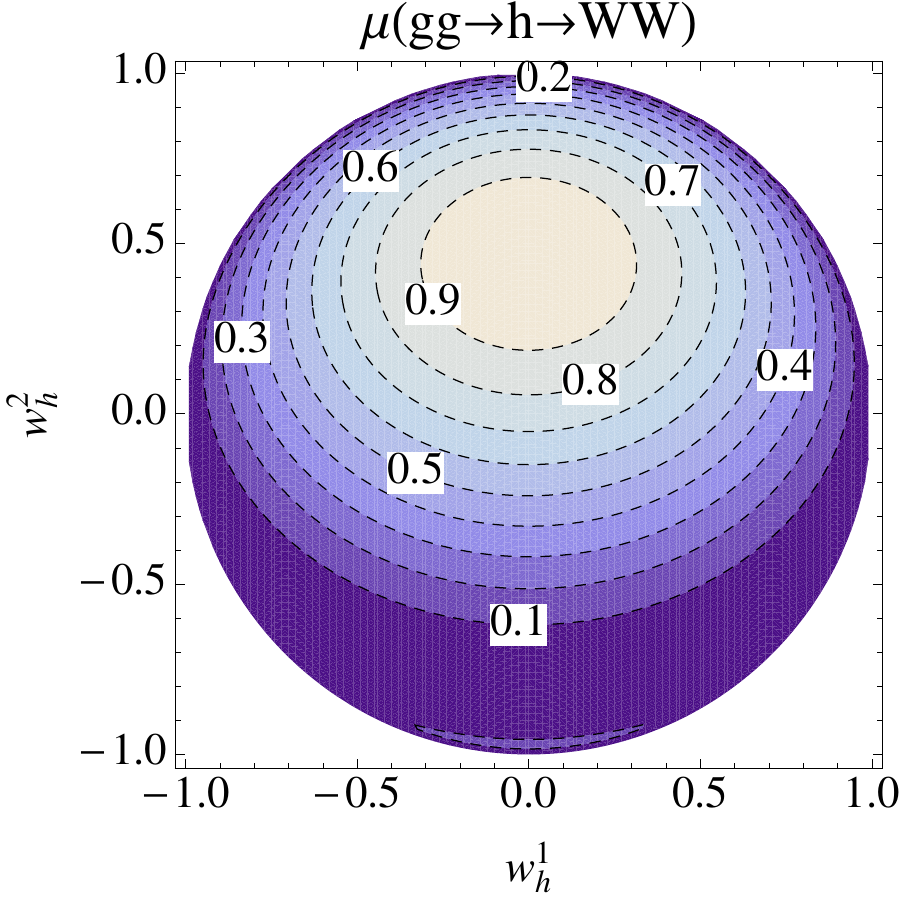}
\end{center}
\end{minipage}
\begin{minipage}{0.24\hsize}
\begin{center}
 \includegraphics[width=\hsize, angle=0]{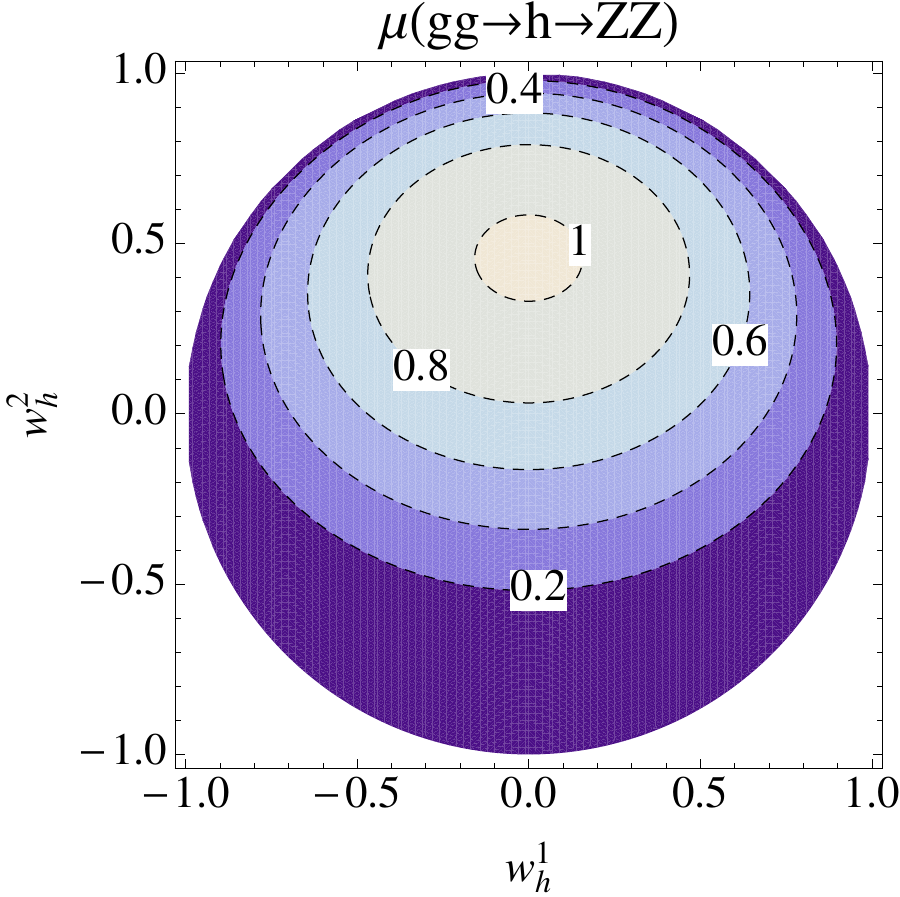}
\end{center}
\end{minipage}
\end{tabular}
 \caption{Signal strengths on the ($w_{h}^{1}$, $w_{h}^2$)-plane. 
 The numbers on the dashed lines are the signal strengths.
 The parameter choices here are $v_3 = 200$~GeV, $m_{W'} = 2500$~GeV, and $r =
 0.1$.
}
\label{fig:mu(vh1,vh2)_200_2500_0.1.eps}
 \end{figure}

\section{conclusion}
In this paper we constructed a model in which $W$/$Z$ and the Higgs
boson are partially composite, and explore the effects of new particles
on the electroweak precision measurements and the signal strength of the
Higgs boson.  We showed that the constraints on $W'$ and $Z'$ from the
electroweak precision measurements and direct searches of them push their
lower mass bounds a few TeV.

In the model we consider, one can take the decoupling limit where $v_1 =
0$ or $v_2 = 0$. In this limit, $W'/Z'$ decouple from the Standard Model
particles.
The vacuum close to such points can naturally be realized since it is
controlled by a soft breaking term of an axial U(1) symmetry, $\kappa$,
in the potential.
In such a vacuum, e.g., $r \equiv v_2/v_1 \gtrsim 5$ or $r \lesssim
0.2$, we find that a relatively small $g_1$, in which perturbative
calculation is valid, is consistent with the electroweak precision
tests. On the other hand, the searches for $W'/Z'$ at the LHC
experiments become important for small $g_1$.
The consistencies with these experimental results are telling us
information on what type of dynamics is behind the electroweak symmetry
breaking. For example, $r \neq 1$ implies parity violating theories
unlike QCD-like technicolor models.

We have calculated the signal strength of $gg \to h \to X$ at the LHC,
and found the Higgs boson at 125~GeV can be partially composite by, for
example, 30\%, whereas all other constraints are satisfied.
If there is a significant composite components in the $W/Z$ bosons, the
Higgs fields should also be partially composite to reproduce the signal
strength measured at the LHC. The deviation from the Standard Model
predictions should be visible in future experiments.

\section*{Acknowledgments}
We would like to thank Mitsutoshi Nakamura for discussion.
RK is supported in part by
the Grant-in-Aid for Scientific Research 23740165 of JSPS.  

\appendix
\section{Elimination of Eq.~(\ref{eq:kappa2})}
\label{sec:kappa2}
%
%
%
In general, the Higgs potential contains the following triple Higgs
interaction terms.
\begin{align}
 \kappa'_1 \tr \left( H_1 H_2 H_3^{\dagger} \right)
+ i \kappa'_2 \tr \left( H_1 H_2 \tau^3 H_3^{\dagger} \right)
.
\end{align}
Note that
\begin{align}
\left(
\tr \left( H_1 H_2 H_3^{\dagger} \right)
\right)^{*}
&=
\tr \left( H_1 H_2 H_3^{\dagger} \right)
, \\ 
\left(
i \tr \left( H_1 H_2 \tau^3 H_3^{\dagger} \right)
\right)^{*}
&=
i \tr \left( H_1 H_2 \tau^3 H_3^{\dagger} \right)
.
\end{align}
Hence $\kappa'_1$ and $\kappa'_2$ are real numbers. We can rewrite the
terms as follows.
\begin{align}
&
 \kappa'_1 \tr \left( H_1 H_2 H_3^{\dagger} \right)
+ i \kappa'_2 \tr \left( H_1 H_2 \tau^3 H_3^{\dagger} \right)
\\
=&
\kappa
\tr \left( H_1 
H_2 \exp(i \tau^3 \theta_{\kappa})
H_3^{\dagger} \right)
,
\end{align}
where
\begin{align}
&
\kappa 
=\sqrt{\kappa_1^{\prime 2} + \kappa_2^{\prime 2}}
, \
 \cos \theta_{\kappa}
=
 \frac{\kappa'_1}{\sqrt{\kappa_1^{\prime 2} + \kappa_2^{\prime 2}}}
, \ 
 \sin \theta_{\kappa}
=
 \frac{\kappa'_2}{\sqrt{\kappa_1^{\prime 2} + \kappa_2^{\prime 2}}}
.
\end{align}
By the field redefinition of $H_2$, we can eliminate $\exp(i \tau^3
\theta_{\kappa})$, namely
\begin{align}
 H_2 \exp(i \tau^3 \theta_{\kappa})
\to&
 H_2
.
\end{align}
This redefinition does not change other terms. Hence we can always
eliminate 
$\tr \left( H_1 H_2 \tau^3 H_3^{\dagger} \right)$.

\section{Constraints from electroweak precision measurements}
\label{sec:STdetail}
The explicit expressions of self-energies after heavy states are
integrated out are
\begin{align}
 \Pi_{W_3 W_3}(q^2)
=&
 \frac{1}{g_0^2} q^2
-
\frac{1}{4} (v_1^2 + v_3^2)
-
\frac{1}{4} v_1^2
\frac{g_1^2}{q^2 - g_1^2 (v_1^2 + v_2^2)/4}
\frac{1}{4} v_1^2
, \\ 
 \Pi_{W_3 B}(q^2)
=&
\frac{1}{4} v_3^2
-
\frac{1}{4} v_1^2
\frac{g_1^2}{q^2 - g_1^2 (v_1^2 + v_2^2)/4}
\frac{1}{4} v_2^2
, \\ 
 \Pi_{B B}(q^2)
=&
 \frac{1}{g_2^2} q^2
-
\frac{1}{4} (v_2^2 + v_3^2)
-
\frac{1}{4} v_2^2
\frac{g_1^2}{q^2 - g_1^2 (v_1^2 + v_2^2)/4}
\frac{1}{4} v_2^2
.
\end{align}
We introduce the following short-handed notations:
\begin{align}
 \Pi'(0) =& 
\left.
\frac{d \Pi (q^2)}{d q^2}
\right|_{q^2 =0}
, \\ 
 \Pi''(0) =& 
\left.
\frac{d^2 \Pi (q^2)}{d (q^2)^2}
\right|_{q^2 =0}
.
\end{align}
Then we find
\begin{align}
 \Pi_{W_3 W_3}(0)
=&
 -\frac{1}{4} (v_1^2 + v_3^2)
 +
\frac{1}{4}\frac{v_1^4}{v_1^2 + v_2^2}
, \\ 
 \Pi_{W_3 W_3}'(0)
=&
 \frac{1}{g_0^2}
 +
\frac{1}{g_1^2}\frac{v_1^4}{(v_1^2 + v_2^2)^2}
, \\ 
 \Pi_{W_3 W_3}''(0)
=&
8
\frac{1}{g_1^4}\frac{v_1^4}{(v_1^2 + v_2^2)^3}
, \\ 
 \Pi_{W_3 B}(0)
=&
 \frac{1}{4} v_2^2
+
\frac{1}{4}\frac{v_1^2 v_2^2}{v_1^2 + v_2^2}
, \\ 
 \Pi_{W_3 B}'(0)
=&
\frac{1}{g_1^2}\frac{v_1^2 v_2^2}{(v_1^2 + v_2^2)^2}
, \\ 
 \Pi_{W_3 B}''(0)
=&
8
\frac{1}{g_1^4}\frac{v_1^2 v_2^2}{(v_1^2 + v_2^2)^3}
, \\ 
 \Pi_{B B}(0)
=&
 -\frac{1}{4} (v_2^2 + v_3^2)
 +
\frac{1}{4}\frac{v_2^4}{v_1^2 + v_2^2}
, \\ 
 \Pi_{BB}'(0)
=&
 \frac{1}{g_2^2}
 +
\frac{1}{g_1^2}\frac{v_2^4}{(v_1^2 + v_2^2)^2}
, \\ 
 \Pi_{BB}''(0)
=&
8
\frac{1}{g_1^4}\frac{v_2^4}{(v_1^2 + v_2^2)^3}
.
\end{align}
From these results, we find
\begin{align}
 g^{-2} \hat{S}
=&
 \Pi_{W_3 B}'(0)
=
\frac{1}{g_1^2} \frac{v_1^2 v_2^2}{(v_1^2 + v_2^2)^2}
, \\ 
2 g^{-2} m_{W}^{-2} W
=&
 \Pi_{W_3 W_3}''(0)
=
8 \frac{1}{g_1^4} \frac{v_1^4}{(v_1^2 + v_2^2)^3}
, \\ 
2 g'^{-2} m_{W}^{-2} Y
=&
 \Pi_{B B}''(0)
=
8 \frac{1}{g_1^4} \frac{v_2^4}{(v_1^2 + v_2^2)^3}
, \\ 
g^{-2} 
=&
 \Pi_{W_1 W_1}'(0)
=
 \Pi_{W_3 W_3}'(0)
=
\frac{1}{g_0^2}
+
\frac{1}{g_1^2}
\frac{v_1^4}{(v_1^2 + v_2^2)^2}
, \\ 
g'^{-2} 
=&
 \Pi_{B B}'(0)
=
\frac{1}{g_2^2}
+
\frac{1}{g_1^2}
\frac{v_2^4}{(v_1^2 + v_2^2)^2}
.
\end{align}
Final results are given in Sec.~\ref{sec:ST}

\section{ Theoretical constraints on parameters in gauge sector}
\label{sec:gaugeDetails}

Note that the trace of a mass matrix gives the sum of the masses and the
determinant of a mass matrix gives the multiple of the masses, so
\begin{align}
 m_{W'}^2 + m_{W}^2
=&
 \frac{1}{4}
\left(
g_0^2 (v_1^2 + v_3^2)
+
g_1^2 (v_1^2 + v_2^2)
\right)
, \\ 
m_{W}^2 m_{W'}^2
=&
\frac{1}{16}
g_0^2 g_1^2
\left(
(v_1^2 + v_3^2)(v_1^2 + v_2^2) - v_1^4
\right)
.
\end{align}
Using these relation, we find
\begin{align}
 (m_{W'}^2 - m_{W}^2)^2 
=&
 (m_{W'}^2 + m_{W}^2)^2 - 4 m_{W'}^2 m_{W}^2
\\ 
=&
\frac{1}{16}
\left(
\left(
g_0^2 ( v_1^2 + v_3^2)
-
g_1^2 ( v_1^2 + v_2^2)
\right)^2
+
4 v_1^4 g_0^2 g_1^2
\right)
\\ 
\geq&
\frac{1}{4}
v_1^4 g_0^2 g_1^2
\\ 
=&
4 v_1^4
\frac{m_{W}^2 m_{W'}^2}{(v_1^2 + v_3^2)(v_1^2 + v_2^2) - v_1^4}
\\ 
=&
4 m_{W}^2 m_{W'}^2
\frac{1}{r^2}
\left(
1
-
\frac{v_3^2}{v^2}
\right)
.
\end{align}
Now we derived Eq.~(\ref{eq:lower_r}).

In $g_{1} \gg g_{0,2}$ region, we can easily express $v_1$ and $v_2$ as
functions of ($m_{W'}$, $g_1$, $v$, $v_3$) by using Eqs.~(\ref{eq:v})
and (\ref{eq:W'mass}):
\begin{align}
 v_{1,2}^2
=&
 \frac{2}{g_1^2}
\left(
m_{W'}^2
\mp
\sqrt{
m_{W'}^2
\left(
m_{W'}^2
-
g_1^2
(v^2 - v_3^2)
\right)
}
\right)
.
\end{align} 
Here we keep only the leading term in Eq.~(\ref{eq:W'mass}). Since
$v_{1,2}$ is real, $\sqrt{\cdots}$ part should be positive and less than
$m_{W'}^2$, then
\begin{align}
0
\leq
m_{W'}^2
-
g_1^2
(v^2 - v_3^2)
<
m_{W'}^2
.
\end{align} 
From Eq.~(\ref{eq:v}), we find $v^2 > v_3^2$, so the above expression is
reduced to
\begin{align}
g_1^2
(v^2 - v_3^2)
\leq
m_{W'}^2
\end{align}
Now we derived Eq.~(\ref{eq:upper_g1}.)

\end{document}